\documentclass[reprint,amsmath,amssymb,nofootinbib,aps,prd]{revtex4-1}

\usepackage{graphicx}% Include figure files
\usepackage{dcolumn}% Align table columns on decimal point
\usepackage{bm}% bold math

\usepackage{hhline}
\usepackage[american]{babel}
\usepackage{booktabs}
\usepackage{dcolumn}  % Align table columns on decimal point
\usepackage[T1]{fontenc}  % Font encoding that enables acents in the PDF
\usepackage[colorlinks]{hyperref}  % Add hypertext capabilities
\usepackage[utf8]{inputenc}  % Accept the Western European characters set
\usepackage{multirow}
\usepackage{times}  % Fonts: Times (text) and Txfonts (math)
\usepackage{url}
\usepackage[dvipsnames]{xcolor}
\usepackage{float}
\usepackage{resizegather}
\usepackage{appendix}

\newcolumntype{d}[1]{D{.}{.}{#1}}
\newcolumntype{v}[1]{D{,}{,\ }{#1}}
\begin{document}
	
	%\preprint{APS/123-QED}
	
	\title{Higher-order extension of Starobinsky inflation: Initial conditions, slow-roll regime, and reheating phase}
	
	\author{G. Rodrigues-da-Silva}
	\email{gesiel.neto.090@ufrn.edu.br}
	\affiliation{Departamento de Física, Universidade Federal do Rio Grande do Norte,\\
	Campus Universitário, s/n - Lagoa Nova, CEP 59072-970, Natal, Rio Grande do Norte, Brazil}
	
	\author{J. Bezerra-Sobrinho}
	\email{jeremias.bs@gmail.com}
	\affiliation{Escola de Ciências e Tecnologia, 	Universidade Federal do Rio Grande do Norte,\\
		Campus Universitário, s/n - Lagoa Nova, CEP 59072-970, Natal, Rio Grande do Norte, Brazil}
	
	\author{L. G. Medeiros}
	\email{leo.medeiros@ufrn.br}
	\affiliation{Escola de Ciências e Tecnologia, 	Universidade Federal do Rio Grande do Norte,\\
	Campus Universitário, s/n - Lagoa Nova, CEP 59072-970, Natal, Rio Grande do Norte, Brazil}

	\date{\today}% It is always \today, today,

\begin{abstract}

The most current observational data corroborate the Starobinsky model as one
of the strongest candidates in the description of an inflationary regime.
Motivated by such success, extensions of the Starobinsky model have been
increasingly recurrent in the literature. The theoretical justification for
this is well grounded: higher-order gravities arise in high-energy physics in
the search for the ultraviolet completeness of general relativity. In this
paper, we propose to investigate the inflation due to the extension of the
Starobinsky model characterized by the inclusion of the $R^{3}$ term. We make
a complete analysis of the potential and phase space of the model, where we
observe the existence of three regions with distinct dynamics for the scalar
field. We can establish restrictive limits for the number of $e$-folds
through a study of the reheating and by considering the usual couplings of the
standard matter fields and gravity. Thereby, we duly confront our model with
the observational data from Planck, BICEP3/Keck, and BAO. Finally, we discuss
how the inclusion of the cubic term restricts the initial conditions
necessary for the occurrence of a physical inflation.

\end{abstract}

\maketitle

\section{Introduction} \label{Introduction}

The idea of a quasi de Sitter type accelerated expansion regime through which the early universe experimented is, on the one hand, a
necessity for the solution of a series of problems that arise in a decelerated
Friedmann universe \cite{Guth:1980zm} and it is, on the other hand, a paradigm
since it has the support of the most current observations from the Planck
satellite \cite{Akrami:2018odb}. Inflationary cosmology emerged with the aim
of proposing a solution to problems such as the horizon, the flatness, and the
magnetic monopoles \cite{Linde:1981mu}. However, over the years, the main
motivation for building an inflationary model has become to provide
a causal mechanism capable of describing the origin of the primordial
inhomogeneities which will be responsible for producing the large-scale structures of the
universe \cite{mukhanov,weinberg}. We have lived in a golden moment in
cosmology mainly because in the last two decades, it has become a branch of
science capable of being duly confronted with observational data.

A consistent inflationary model must last long enough to deal with problems
like the ones mentioned above and must also end up, through the graceful exit,
giving way to Friedmann's decelerated universe, in order to preserve the
successful predictions of the standard model of cosmology. There are numerous
models proposed in the literature suggesting the existence of a scalar field,
or multiple ones, driving inflation \cite{Linde:1993cn,Wands:2007bd,MARTIN201475}. Some of these models explore the idea of scalar fields,
arising from some fundamental interaction, evolving in curved spacetimes
described by general relativity (GR) \cite{Linde:1983gd,Freese:1990rb,Bezrukov:2007ep}, and some others are based on GR
extensions, through the inclusion of higher-order curvature terms in the
Einstein-Hilbert (EH) action \cite{doi.org/10.1016/0370-2693(80)90670-X,PhysRevD.43.2510,PhysRevD.91.083529,Ivanov:2016hcm,Salvio:2017xul,Cuzinatto:2018vjt}.

Higher-order gravities are motivated by high-energy physics in the search for
the ultraviolet (UV) completeness of GR. As it is well known, GR is a
nonrenormalizable theory, and it is not possible to conventionally quantize
it. However, the inclusion of such higher-order curvature terms --- involving
functions of the curvature scalar $R$, contractions of the Ricci tensor
$R_{\mu\nu}$ and/or Riemann $R_{\kappa\lambda\mu\nu}$, as well as their
derivatives --- contribute to its renormalizability
\cite{doi.org/10.1063/1.1724264}. K. S. Stelle, for example, obtained a
renormalizable system due to the inclusion of quadratic curvature correction
terms in EH action \cite{PhysRevD.16.953}. In this sense, GR could be seen as
an effective low-energy theory that requires higher-order curvature
corrections as we increase the energy scale \cite{AIHPA_1974__20_1_69_0}. The
inclusion of these terms is often accompanied by the problem of
introducing ghost-type instabilities, which quantumly manifests itself with
the loss of unitarity and, classically, with the absence of a lower limit for
the Hamiltonian of the system \cite{Sbisa:2014pzo}. However, this is not a
problem that arises in $f(R)$ type extensions, known to be ghost-free
\cite{RevModPhys.82.451}.

The Starobinsky model \cite{doi.org/10.1016/0370-2693(80)90670-X}, which is
characterized by including the quadratic term $R^{2}$ in EH action, and
consequently introducing only one additional parameter, is one of the
strongest inflationary candidates, best fitting to the most current
observational data\footnote{Along with Higgs inflation \cite{Bezrukov:2007ep}, since they have the same potential.} \cite{Akrami:2018odb}. By proposing an
extension to GR, an inflationary regime can be obtained, for a certain range
of the theory parameters, without the need of the extra fields used in part of
the inflation mechanisms. This is due to the equivalence between higher-order
gravities, whether $f\left(  R\right)  $ or $f\left(  R,\square^{k}R\right)
$, and the scalar-tensor gravity theories \cite{Gottlober:1989ww,RevModPhys.82.451,Capozziello:2011et,Cuzinatto:2016ehv}. In
these cases, through a conformal transformation, we go from the original frame
with equations for the metric of order $2l+2$ to the Jordan or Einstein frame
with $l$ scalar fields responsible for driving
inflation \cite{Amendola:1993bg}. With this scenario, researches considering extensions to the
Starobinsky model has been increasingly recurrent in the literature
\cite{Iihoshi:2010pf,Huang:2013hsb,Asaka:2015vza,Pi:2017gih,Castellanos:2018dub,Cuzinatto:2018vjt,Cheong:2020rao,Cano:2020oaa,Chojnacki:2021fag}.

In this paper, we propose the generalization of Starobinsky inflation due to
the inclusion of the cubic term $R^{3}$ in the gravitational action. So we can
write the action that describes this model as%
\begin{equation}
S\left(  g_{\mu\nu}\right)  =\frac{M_{Pl}^{2}}{2}\int d^{4}x\sqrt{-g}\left(
R+\frac{1}{2\kappa_{0}}R^{2}+\frac{\alpha_{0}}{3\kappa_{0}^{2}}R^{3}\right)  ,
\label{Acao do Trabalho}%
\end{equation}
where $M_{Pl}$ is the reduced Planck mass, so that $M_{Pl}^{2}\equiv\left(
8\pi G\right)  ^{-1}$, $\kappa_{0}$ has squared mass units and the parameter
$\alpha_{0}$ is a dimensionless quantity. For consistency with the Starobinsky
model, we assume $\kappa_{0}>0$. We can view the action
(\ref{Acao do Trabalho}) as a classical theory of gravity containing
higher-order energy corrections to GR.

The paper is structured as follows. In Sec. \ref{Sec:field-equations}, we
present the action (\ref{Acao do Trabalho}) in the context of a scalar-tensor
theory in the Einstein frame and its associated potential. In Sec.
\ref{sec - pot}, we make a complete study of the potential and phase space of
the model. We observe the existence of three regions with distinct dynamics
for the scalar field, and in addition, we qualitatively discuss the influence
that the cubic term has on the restriction of the initial conditions
capable of leading the system to a consistent inflationary regime (physical
inflation). In Sec. \ref{Sec:inflationary-dynamics}, we describe the entire
dynamics of the scalar field, focusing (i) on the region where the slow-roll
regime occurs and (ii) in the reheating region, where we develop a study
that allows us to obtain expressions for the reheating number of $e$-folds and
temperature valid for a wide range of models, and then we particularize for
our case. In Sec. \ref{sec - obs}, by using our results on the reheating
period and considering the minimum hypothesis of the usual couplings of the
standard matter fields with gravity, we can obtain restrictive limits for the
inflation\ number of $e$-folds. In its turn, with the interval obtained for the
number of $e$-folds, we confront our model with the observations of Planck,
BICEP3/Keck and BAO \cite{Akrami:2018odb,BICEPKeck:2021gln}, by
establishing the constraints on the theoretical values of scalar spectral
index $n_{s}$ and the tensor-to-scalar ratio $r$. Finally, we discuss the
fine-tuning problem of the initial conditions that the $R^{3}$ term introduces for the occurrence of a physical inflation, as well as the conditions that lead to an eternal inflation regime. In Sec.
\ref{sec:final-comments}, we make some final comments.

\section{Field equations} \label{Sec:field-equations}

The action (\ref{Acao do Trabalho}) can be conveniently rewritten as a
scalar-tensor theory in the Einstein frame, in which we have a description
through an auxiliary scalar field $\chi$ minimally coupled with gravitation.
By following the steps presented in Appendix \ref{Ap1}, we write%
\begin{equation}
S\left(  \bar{g}_{\mu\nu},\chi\right)  =\frac{M_{Pl}^{2}}{2}\int d^{4}%
x\sqrt{-\bar{g}}\left[  \bar{R}-3\left(  \frac{1}{2}\bar{\partial}^{\rho}%
\chi\bar{\partial}_{\rho}\chi+V\left(  \chi\right)  \right)  \right]
,\label{Acao Frame de Einstein Adim}%
\end{equation}
whose associated potential is%
\begin{align}
V\left(  \chi\right)   &  =\frac{\kappa_{0}}{72\alpha_{0}^{2}}e^{-2\chi
}\left(  1-\sqrt{1-4\alpha_{0}\left(  1-e^{\chi}\right)  }\right)
\times\nonumber\\
&  \times\left[  -1+8\alpha_{0}\left(  1-e^{\chi}\right)  +\sqrt{1-4\alpha
	_{0}\left(  1-e^{\chi}\right)  }\right]  . \label{Potencial V(Chi)}%
\end{align}
The barred quantities are defined from the metric as $\bar{g}_{\mu\nu}%
=e^{\chi}g_{\mu\nu}$ and the dimensionless field $\chi$ is defined as%
\begin{equation}
e^{\chi}=1+\frac{R}{\kappa_{0}}+\alpha_{0}\left(  \frac{R}{\kappa_{0}}\right)
^{2}\text{.} \label{Chi funcao R}%
\end{equation}

\subparagraph{Addendum}

In order to recover the usual notation and dimensions of the scalar field and
the potential, we must do%
\begin{equation}
\chi=\sqrt{\frac{2}{3}}\frac{\phi}{M_{Pl}}\text{ \ and \ }\bar{V}\left(
\phi\right)  =\frac{3M_{Pl}^{2}}{2}V\left(  \chi\right)  .
\label{campo escalar dimensional}%
\end{equation}
In this case, the action (\ref{Acao Frame de Einstein Adim}) is rewritten as%
\[
S\left(  \bar{g}_{\mu\nu},\phi\right)  =\int d^{4}x\sqrt{-\bar{g}}\left(
\frac{M_{Pl}^{2}}{2}\bar{R}-\frac{1}{2}\bar{\partial}^{\rho}\phi\bar{\partial
}_{\rho}\phi-\bar{V}\left(  \phi\right)  \right)  .
\]

\bigskip

The action (\ref{Acao Frame de Einstein Adim}) gives us two field equations:
one for the metric $\bar{g}_{\mu\nu}$ and one for the scalar field $\chi$. By
taking the variation concerning the metric $\bar{g}_{\mu\nu}$, we obtain%
\begin{equation}
\bar{R}_{\mu\nu}-\frac{1}{2}\bar{g}_{\mu\nu}\bar{R}=\frac{1}{M_{Pl}^{2}}%
\bar{T}_{\mu\nu}^{\left(  \text{eff}\right)  }, \label{eq:campo-metrica}%
\end{equation}
where%
\begin{equation}
\frac{1}{M_{Pl}^{2}}\bar{T}_{\mu\nu}^{\left(  \text{eff}\right)  }=\frac{3}%
{2}\left[  \bar{\partial}_{\mu}\chi\bar{\partial}_{\nu}\chi-\bar{g}_{\mu\nu
}\left(  \frac{1}{2}\bar{\partial}^{\rho}\chi\bar{\partial}_{\rho}%
\chi+V\left(  \chi\right)  \right)  \right]  . \label{T_munu efetivo}%
\end{equation}
In turn, the variation concerning the scalar field $\chi$ results in%
\begin{equation}
\bar{\square}\chi-V^{\prime}\left(  \chi\right)  =0, \label{eq:eq-campo-chi}%
\end{equation}
where the box $\bar{\square}\equiv\bar{\nabla}^{\rho}\bar{\nabla}_{\rho}$
represents the covariant d'Alembertan operator and the prime represents the
derivative with respect to $\chi$. By calculating $V^{\prime}\left(
\chi\right)  $ explicitly, we obtain%
\begin{align}
V^{\prime}\left(  \chi\right)   &  =\frac{\kappa_{0}}{36\alpha_{0}^{2}%
}e^{-2\chi}\left(  -1+\sqrt{1-4\alpha_{0}\left(  1-e^{\chi}\right)  }\right)
\times\nonumber\\
&  \times\left[  -1+2\alpha_{0}\left(  4-e^{\chi}\right)  +\sqrt{1-4\alpha
	_{0}\left(  1-e^{\chi}\right)  }\right]  . \label{V linha}%
\end{align}

\section{Potential and phase space in the Friedmann background}	\label{sec - pot}

We begin this section by writing the field equations in the Friedmann
background. Considering the flat Friedmann-Lema\^{\i}tre-Robertson-Walker
(FLRW) metric%
\begin{equation}
ds^{2}=-dt^{2}+a^{2}\left(  t\right)  \left(  dx^{2}+dy^{2}+dz^{2}\right)  ,
\label{eq:metrica-friedmann}%
\end{equation}
we obtain%
\begin{equation}
H^{2}=\frac{1}{2}\left(  \frac{1}{2}\dot{\chi}^{2}+V\left(  \chi\right)
\right)  , \label{H2 Sta plus R3}%
\end{equation}%
\begin{equation}
\dot{H}=-\frac{3}{2}\left(  \frac{1}{2}\dot{\chi}^{2}\right)  ,
\label{Hpto Sta plus R3}%
\end{equation}
and%
\begin{equation}
\ddot{\chi}+3H\dot{\chi}+V^{\prime}\left(  \chi\right)  =0,
\label{Chi Sta plus R3}%
\end{equation}
where the dot represents time derivative and $H=\dot{a}/a$ is the Hubble
parameter. The potential and its derivative are given by Eqs.
(\ref{Potencial V(Chi)}) and (\ref{V linha}), with $\chi=\chi\left(  t\right)
$.

\begin{figure}[h] 
	
\centering
\includegraphics[scale=0.5]{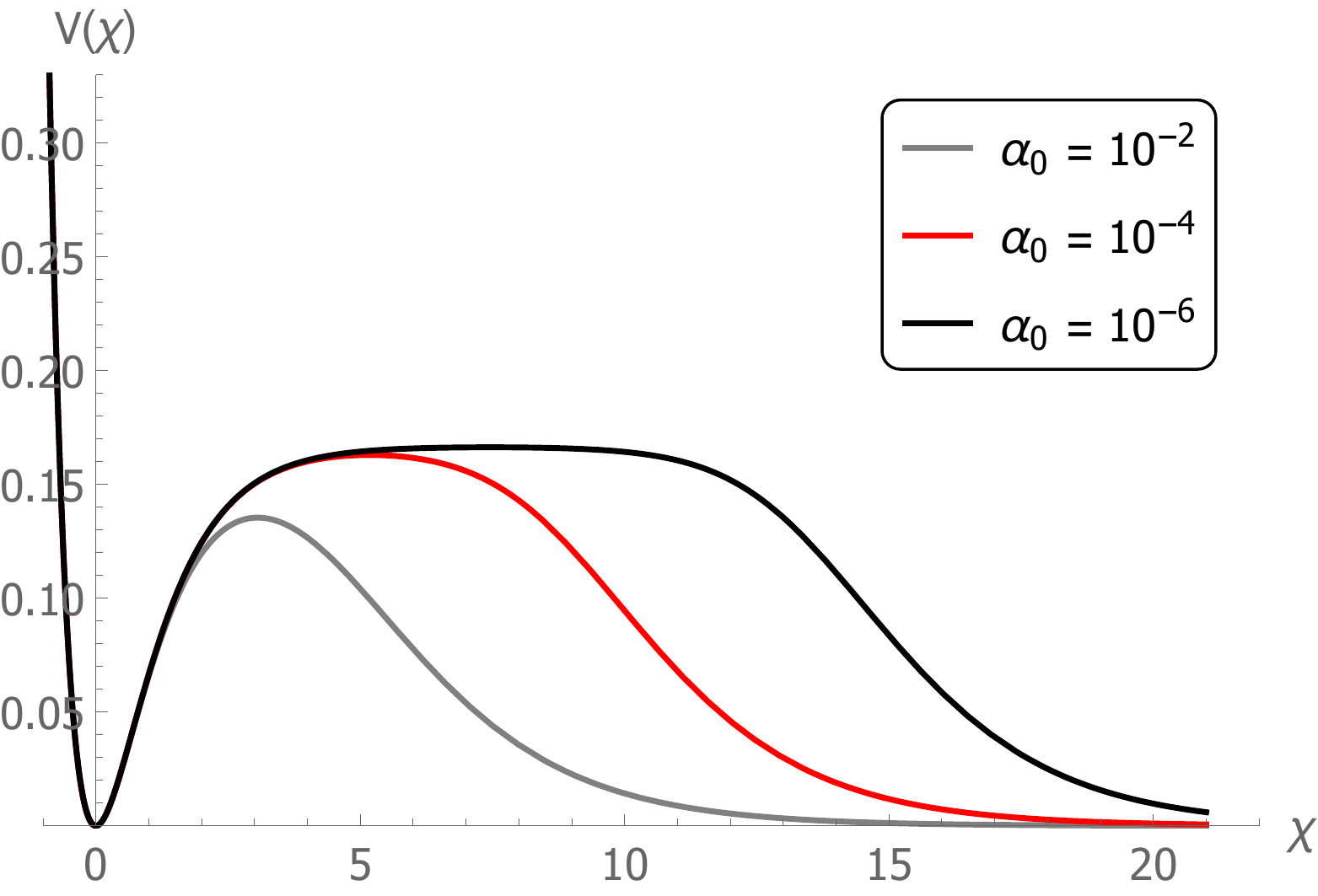}
\caption{\label{fig:potencial} Potential $V\left(  \chi\right)  $ as a function of $\chi$,
	normalized to $\kappa_{0}=1$. The maximum values are located at $\chi
	_{c}\simeq3.06$, $5.18$ and $7.46$ for $\alpha_{0}=10^{-2}$, $10^{-4}$ and
	$10^{-6}$, respectively.}
\end{figure}

The first steps to be analyzed are the
characteristics of the potential $V\left(  \chi\right)  $. The structure of
the potential depends on the parameter $\alpha_{0}$. For $V\left(
\chi\right)  $ to be well defined for all real numbers, it is necessary
that\footnote{Throughout this paper, we will assume that $\alpha_{0}$ is
	contained in this range.}%
\begin{equation}
0\leq\alpha_{0}\leq\frac{1}{4}. \label{Restricao em alpha zero}%
\end{equation}
The lower (upper) limit of $\alpha_{0}$ makes $V\left(  \chi\right)  $ a real
function for any value of $\chi$ greater (less) than zero. Furthermore,
$\alpha_{0}\geq0$ guarantees the stability of the gravitational model during
the entire inflationary regime\footnote{The $f\left(  R\right)  $ models are
	stable whenever $f^{\prime}\left(  R\right)  >0$ and $f^{\prime\prime}\left(
	R\right)  \geq0$ \cite{RevModPhys.82.451}. For our case%
	\[
	f\left(  R\right)  =R+\frac{1}{2\kappa_{0}}R^{2}+\frac{\alpha_{0}}{3\kappa
		_{0}^{2}}R^{3},
	\]
	where
	\[
	R=6\left(  \dot{H}+2H^{2}\right)  =-\frac{3}{2}\dot{\chi}^{2}+6V\left(
	\chi\right)  .
	\]
	During the inflationary regime, when $\dot{\chi}^{2}\ll V\left(  \chi\right)
	$ and $\kappa_{0}>0$, the stability condition implies $\alpha_{0}\geq0$.}.
Within the range given in Eq. (\ref{Restricao em alpha zero}), excepting
$\alpha_{0}=0$, the potential has two critical points, namely,%
\begin{equation}
\chi_{0}=0\text{ \ }\rightarrow\text{\ \ (minimum point),} \label{Chi 0}%
\end{equation}%
\begin{equation}
\chi_{c}=\ln\left(  4+\sqrt{\frac{3}{\alpha_{0}}}\right)  \text{
	\ }\rightarrow\text{ \ (maximum point).} \label{Chi c}%
\end{equation}
The behavior of the potential for different values of $\alpha_{0}$ is shown in
Fig. \ref{fig:potencial}.

The plateau that appears in Fig. \ref{fig:potencial} gives us an indication that a slow-roll
type inflationary regime can be achieved. We observe that the smaller the
magnitude of $\alpha_{0}$, the more the point $\chi_{c}$ moves to the right
and the larger the plateau becomes. So, at the limit of $\alpha
_{0}\rightarrow0$, the plateau becomes infinite, and we recover the
Starobinsky inflation. The graphic structure in Fig. \ref{fig:potencial} also seems to
indicate that for $\chi<\chi_{c}$ we have a slow-roll regime to the left,
while for $\chi>\chi_{c}$ we have a slow-roll regime to the right. In order to
better understand this dynamic, we must carry out a study of the phase space
of this model.

The phase space of this model can be analyzed by combining Eqs.
(\ref{H2 Sta plus R3}) and (\ref{Chi Sta plus R3}) into a single ODE whose
independent variable is $\chi$. In this case, we have%
\begin{equation}
\frac{d\dot{\chi}}{d\chi}=\frac{-3\sqrt{\frac{1}{4}\dot{\chi}^{2}+\frac{1}%
		{2}V\left(  \chi\right)  }\dot{\chi}-V^{\prime}\left(  \chi\right)  }%
{\dot{\chi}}, \label{Phase space equation}%
\end{equation}
where $V\left(  \chi\right)  $ and $V^{\prime}\left(  \chi\right)  $ are given
by Eqs. (\ref{Potencial V(Chi)}) and (\ref{V linha}), respectively. The
analysis of Eq. (\ref{Phase space equation}) as a parametric autonomous system
shows that phase space has two critical points located at $\left(  \chi
,\dot{\chi}\right)  _{0}=\left(  0,0\right)  $ and $\left(  \chi,\dot{\chi
}\right)  _{c}=\left(  \chi_{c},0\right)  $ where $\chi_{c}$ is given in Eq.
(\ref{Chi c}). Furthermore, the linearized analysis of this system around the
critical points shows that $\left(  \chi_{c},0\right)  $ is a unstable saddle
point, while for $\left(  0,0\right)  $, such characterization is
inconclusive\footnote{In the linearized description in the vicinity of the
	critical points of the system, when obtaining associated pure imaginary roots,
	as occurs in $\left(  \chi,\dot{\chi}\right)  _{0}=\left(  0,0\right)  $, its
	nature is uncertain: it may be a center or a spiral point. Next, we show
	through numerical methods that the critical point $\left(  \chi,\dot{\chi
	}\right)  _{0}$ is an attractor spiral point.}.

The complete description of the phase space is developed by a numerical
analysis of Eq. (\ref{Phase space equation}).

\begin{figure}[h]
	
	\centering
	\includegraphics[scale=0.31]{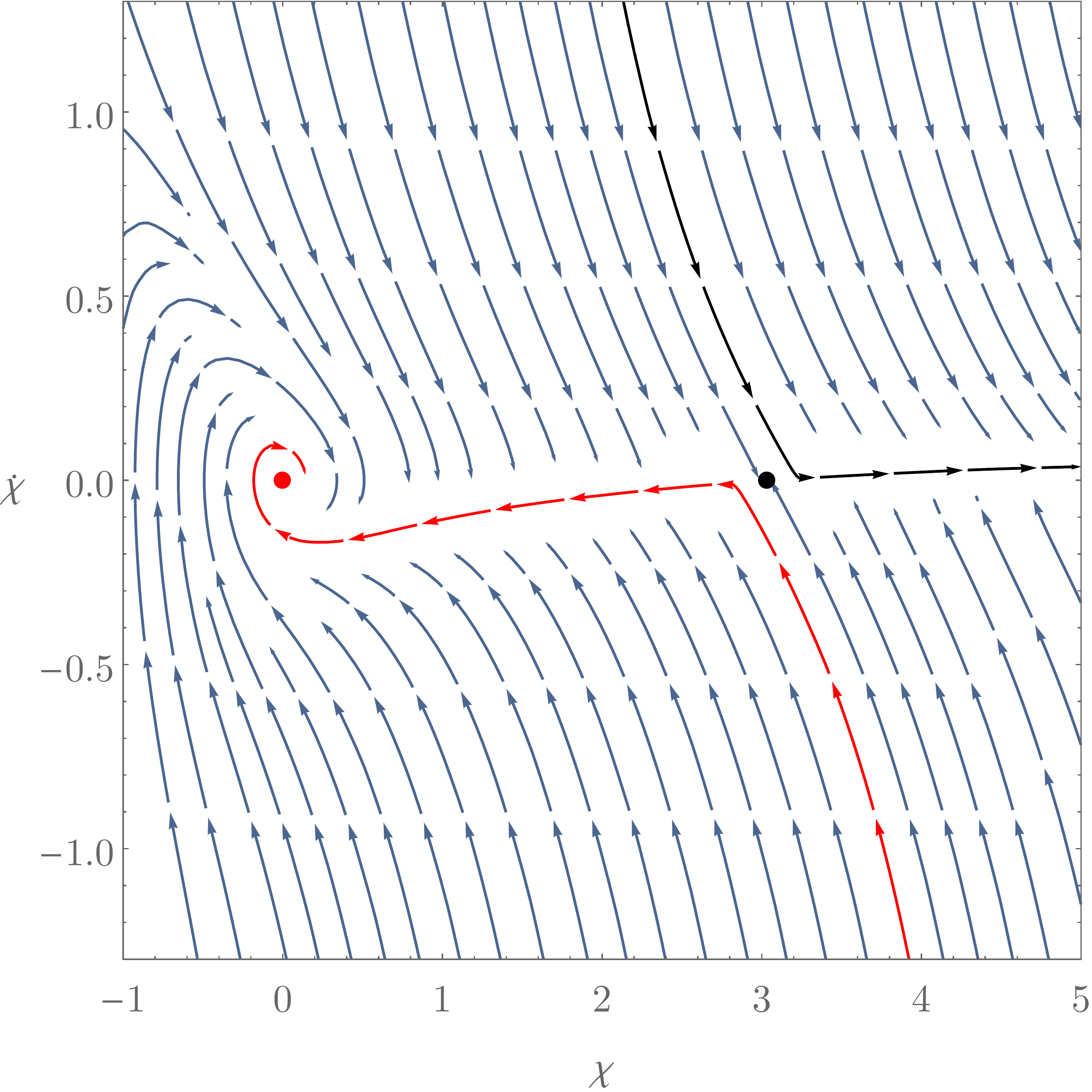}
	\caption{\label{fig:phase-space} Phase space $\left(  \chi,\dot{\chi}\right)  $ considering
		$\kappa_{0}=1$ and $\alpha_{0}=10^{-2}$. The red and black points correspond
		to the critical points $\left(  0,0\right)  $ and $\left(  \chi_{c},0\right)
		$ with $\chi_{c}=3.06$, respectively. In its turn, the red and black lines
		represent two trajectories that travel in opposite directions concerning the
		saddle point $\left(  \chi_{c},0\right)  $.}
\end{figure}

We can make several conclusions about the phase space of the system from the
Fig. \ref{fig:phase-space}. We observe that a large number of initial conditions
($\chi\gtrsim2$ and any $\dot{\chi}$) are conducted to the attractor line
existing in the vicinity of $\dot{\chi}\approx0$. Furthermore, in the
subregion of this attractor line that corresponds to the potential plateau in
Fig. \ref{fig:potencial}, we have a slow-roll regime where $\dot{\chi}\ll1$ and $\chi$ is at
least of the order of the unity. As expected, the slow-roll regime leads to an
almost exponential expansion that characterizes inflation. That can be
explicitly identified when comparing Eq. (\ref{H2 Sta plus R3}) with Eq.
(\ref{Hpto Sta plus R3}). In fact, in the plateau region we have $\dot{\chi
}^{2}\ll V\left(  \chi\right)  $ (compare the numerical values of Figs. \ref{fig:potencial}
and \ref{fig:phase-space} in that region), so%
\[
\left\vert \frac{\dot{H}}{H^{2}}\right\vert \simeq\frac{3\dot{\chi}^{2}%
}{2V\left(  \chi\right)  }\ll1.
\]
The above condition is the necessary one for a slow-roll type inflationary
regime to take place.

A second important finding is that the critical point $\left(  \chi
_{c},0\right)  $ divides the attractor line into two distinct regions. If we
are to the left side of this point, the slow-roll takes place to the left (red
line in Fig. \ref{fig:phase-space}) and inflation unfolds in the usual way. In this case, the
scalar field $\chi$ moves toward the critical point $\left(  0,0\right)  $
and, when approaching it, it spirals continuously around the origin. This
stage corresponds to the phase of coherent oscillations that occurs at the
beginning of the reheating process. In turn, if we are to the right side of
the point $\left(  \chi_{c},0\right)  $, the slow-roll occurs to the right
(black line in Fig. \ref{fig:phase-space}) and the scalar field $\chi$ grows indefinitely. In
that second region, both $\dot{\chi}$ and $V\left(  \chi\right)  $ approach
zero when $\chi\gg1$.

From the discussions related to the potential $V\left(  \chi\right)  $ (Fig.
\ref{fig:potencial}) and the phase space (Fig. \ref{fig:phase-space}), we recognize three distinct regions
associated with the dynamics of the scalar field $\chi$: an asymptotic region,
where $V\left(  \chi\right)  \rightarrow0$ for $\chi\gg1$; a plateau region,
where the slow-roll type inflationary regime occurs; and a region of
oscillations around the origin of the potential that characterizes the end of
inflation. In the next section, we analyze each of these regions in a more
detailed way.

In Ref. \cite{Ivanov:2021chn}, the
authors discuss the Starobinsky$+R^{3}$ and Starobinsky$+R^{4}$ models. It
is shown that there is quite similarity in the nature of both models.
Qualitatively, this is due to the similarity between the potentials of the
models: both have a maximum point with finite $\chi _{c}>0$, in the vicinity of which
there is a plateau region, where a slow-roll inflationary regime may occur.
Through our analysis via phase space, we conclude that, in a classical context, the Starobinsky$+R^{4}
$ model has a dynamic that also restricts the initial conditions
necessary for the occurrence of a physical inflation. A discussion of the
role played by higher-order $R^{n}$ terms with $n>3$ in extensions to the
Starobinsky model is given in the Sec. \ref{sec:final-comments}.

\section{Inflationary dynamics of the scalar field}	\label{Sec:inflationary-dynamics}

In this section, we describe the evolution of the scalar field $\chi$ in the
three regions of interest: the asymptotic region where $V\left(  \chi\right)
\rightarrow0$ when $\chi\gg1$; the plateau region, where the inflationary
regime occurs; and the region of oscillation around the origin of the
potential that characterizes the end of inflation.

\subsection{Asymptotic region of the potential $V\left(  \chi\right)  $}

We saw in Sec. \ref{sec - pot} that depending on the initial conditions for
the scalar field $\chi$, it evolves to the right in the phase space (see
Fig. \ref{fig:phase-space}). In this case, the scalar field $\chi$ grows indefinitely and
reaches an asymptotic region of the potential, namely, $V\left(  \chi\right)
\rightarrow0$. The purpose of this subsection is to analyze the cosmic
dynamics in this region.

In a sufficiently large region of $\chi$, the potential
(\ref{Potencial V(Chi)}) and its derivative (\ref{V linha}) behave like
\[
V\left(  \chi\right)  \approx\frac{2\kappa_{0}}{9\sqrt{\alpha_{0}}}%
e^{-\frac{1}{2}\chi}\text{ \ and\ \ }V^{\prime}\left(  \chi\right)
\approx-\frac{\kappa_{0}}{9\sqrt{\alpha_{0}}}e^{-\frac{1}{2}\chi
}\text{\ \ for\ \ }\chi\gg1\text{.}%
\]
In this case, the field equation (\ref{Chi Sta plus R3}) can be written as%
\[
\ddot{\chi}+\frac{3}{\sqrt{2}}\sqrt{\frac{1}{2}\dot{\chi}^{2}+\frac
	{2\kappa_{0}}{9\sqrt{\alpha_{0}}}e^{-\frac{1}{2}\chi}}\dot{\chi}-\frac
{\kappa_{0}}{9\sqrt{\alpha_{0}}}e^{-\frac{1}{2}\chi}\approx0.
\]
A convenient ansatz for solutions of this equation is $\dot{\chi}=Ae^{-k\chi}%
$. Substituting this proposed solution in the previous equation, we obtain%
\[
k=\frac{1}{4}\text{ \ and \ }A=\frac{2}{3}\sqrt{\frac{\kappa_{0}}%
	{35\sqrt{\alpha_{0}}}}\Rightarrow\dot{\chi}\approx\frac{2}{3}\sqrt
{\frac{\kappa_{0}}{35\sqrt{\alpha_{0}}}}e^{-\frac{1}{4}\chi}.
\]
By having $\dot{\chi}$ and $V\left(  \chi\right)  $, we can explicitly
determine the parameter of the equation of state\footnote{Note $\rho_{\chi
	}=\frac{1}{2}\dot{\chi}^{2}+V\left(  \chi\right)  $ has a squared mass units
	because by the adopted choice $\chi$ is dimensionless.}%
\begin{equation}
w=\frac{p_{\chi}}{\rho_{\chi}}=\frac{\frac{1}{2}\dot{\chi}^{2}-V\left(
	\chi\right)  }{\frac{1}{2}\dot{\chi}^{2}+V\left(  \chi\right)  }\approx
-\frac{17}{18}\text{ \ for\ \ }\chi\gg1\text{.} \label{eq:asymptotic-w}%
\end{equation}
The result obtained in Eq. (\ref{eq:asymptotic-w}) shows that when the scalar
field rolls to the right side from the plateau of the potential, the parameter
of the equation of state leaves a value of $w\approx-1$ (plateau region) and
moves asymptotically to $w\approx-\left(  17/18\right)  $. In other words, the
inflation dynamics migrates from an almost de Sitter regime to a power-law
type inflation with $a\sim t^{12}$. Therefore, in this region, the accelerated
expansion never ends. This explicitly shows that the existence of the critical
point $\left(  \chi_{c},0\right)  $ limits the initial conditions favorable to
a physical inflation i.e. an inflation that connects to a radiation era type
of a hot big-bang model. In Sec. \ref{sec - probabilidade}, such a
limitation on the initial conditions is further discussed.

\subsection{Slow-roll leading-order inflationary regime} \label{sec - slow roll}

In this subsection, we investigate the slow-roll inflationary regime that
occurs in the plateau region of Fig. \ref{fig:potencial}. In leading-order, Eqs.
(\ref{Chi Sta plus R3}) and (\ref{H2 Sta plus R3}) are approximated by%
\begin{equation}
3H\dot{\chi}\approx-V^{\prime}, \label{Eq Chi aprox}%
\end{equation}%
\begin{equation}
H^{2}\approx\frac{1}{2}V\left(  \chi\right)  , \label{H2 aprox}%
\end{equation}
that when combined they result in%
\begin{equation}
\dot{\chi}\left(  \chi\right)  \approx-\frac{\sqrt{2}V^{\prime}}{3\sqrt{V}}.
\label{Chi pto aprox potencial}%
\end{equation}

The next step is obtaining $V$ and $V^{\prime}$ in a leading order slow-roll
regime. Defining a fundamental slow-roll parameter, namely, $\delta\equiv
e^{-\chi}$,\footnote{A slow-roll regime, where the scalar field $\chi$ lies in
	the vicinity of the plateau, occurs for $\delta\ll1$.} we write the potential
(\ref{Potencial V(Chi)}) as%
\begin{align*}
V\left(  \chi\right)   &  =-\frac{\kappa_{0}}{72\alpha_{0}^{2}}\left(
-\delta+\delta\sqrt{1-4\alpha_{0}\delta^{-1}\left(  \delta-1\right)  }\right)
\times\\
&  \times\left[  -\delta+8\alpha_{0}\left(  \delta-1\right)  +\delta
\sqrt{1-4\alpha_{0}\delta^{-1}\left(  \delta-1\right)  }\right]  .
\end{align*}
Furthermore, we know that the maximum value of the potential consistent with a
physical inflationary regime occurs for $V\left(  \chi_{c}\right)  $, where%
\[
e^{\chi_{c}}=4+\sqrt{\frac{3}{\alpha_{0}}}\Rightarrow\alpha_{0}=\frac
{3}{\left(  e^{\chi_{c}}-4\right)  ^{2}}.
\]
So, in a slow-roll leading-order regime, we have%
\begin{equation}
\alpha_{0}\approx\frac{3}{e^{2\chi_{c}}}\equiv3\delta_{c}^{2}.
\label{def de delta min}%
\end{equation}
In its turn, as the minimum value of $\delta$ consistent with a physical
inflation is $\delta_{c}$, we have%
\begin{equation}
4\alpha_{0}\delta^{-1}\approx12\delta_{c}^{2}\delta^{-1}\lesssim12\delta\ll1.
\label{Condicao alpha 0 delta c}%
\end{equation}
The relation (\ref{Condicao alpha 0 delta c}) shows that the quantity
$4\alpha_{0}\delta^{-1}$ contributes at most in the slow-roll leading-order regime. So, it is worth pointing out that $\alpha_{0}$ is a second-order slow-roll term.

From the previous considerations, we can approximate the potential and its
derivatives by%
\begin{equation}
V\left(  \chi\right)  \approx\frac{\kappa_{0}}{6}\left(  1-2\delta-\frac{2}%
{3}\alpha_{0}\delta^{-1}\right)  ,\label{V aprox}%
\end{equation}%
\begin{equation}
V^{\prime}\left(  \chi\right)  \approx\frac{\kappa_{0}}{3}\left(  \delta
-\frac{1}{3}\alpha_{0}\delta^{-1}\right)  ,\label{V linha aprox}%
\end{equation}%
\begin{equation}
V^{\prime\prime}\left(  \chi\right)  \approx-\frac{\kappa_{0}}{3}\left(
\delta+\frac{1}{3}\alpha_{0}\delta^{-1}\right)  .\label{V 2linhas aprox}%
\end{equation}

Since we have determined approximate $V\left(  \chi\right)  $ and $V^{\prime
}\left(  \chi\right)  $ we can calculate $\dot{\chi}$ in the slow-roll
leading-order. Starting from (\ref{Chi pto aprox potencial}) we have%
\begin{equation}
\dot{\chi}\approx-\frac{\sqrt{2}V^{\prime}}{3\sqrt{V}}\Rightarrow\dot{\chi
}\approx-\frac{2\sqrt{3\kappa_{0}}}{9}\left(  \delta-\delta_{c}^{2}\delta
^{-1}\right)  . \label{Chi pto aprox}%
\end{equation}
That is, in the slow-roll leading-order, we have $\dot{\chi}\sim\delta$ plus a
correction proportional to $\delta_{c}^{2}\delta^{-1}\lesssim\delta$.

With the previous results, we can calculate the slow-roll parameters%
\begin{equation}
\epsilon\equiv-\frac{\dot{H}}{H^{2}}\text{ \ and\ \ }\eta\equiv-\frac{1}%
{H}\frac{\dot{\epsilon}}{\epsilon}. \label{slow roll definitions}%
\end{equation}
By using Eqs. (\ref{Hpto Sta plus R3}), (\ref{H2 aprox}), (\ref{V aprox}) and
(\ref{Chi pto aprox}), we get in the slow-roll leading-order, written in terms
the fundamenal parameter $\delta$%
\begin{equation}
\epsilon\approx\frac{4}{3}\left(  \delta-\delta^{-1}\delta_{c}^{2}\right)
^{2}, \label{epsilon 2}%
\end{equation}%
\begin{equation}
\eta\approx-\frac{8}{3}\left(  \delta+\delta^{-1}\delta_{c}^{2}\right)  .
\label{eta 2}%
\end{equation}
The expressions (\ref{epsilon 2}) and (\ref{eta 2}) make it clear that for
$\delta\ll1$ we have $\epsilon\ll1$ and $\eta\ll1$, and therefore we are in an
almost de Sitter expansion regime.

The next step is calculating the number of $e$-folds $N$. By using Eqs.
(\ref{V aprox}) and (\ref{V linha aprox}) we have
\begin{equation}
N=%
%TCIMACRO{\dint \limits_{t}^{t_{e}}}%
%BeginExpansion
{\displaystyle\int\limits_{t}^{t_{e}}}
%EndExpansion
Hdt\approx\frac{3}{4}%
%TCIMACRO{\dint \limits_{\delta}^{\delta_{e}}}%
%BeginExpansion
{\displaystyle\int\limits_{\delta}^{\delta_{e}}}
%EndExpansion
\frac{1-2\delta-2\delta^{-1}\delta_{c}^{2}}{\delta-\delta^{-1}\delta_{c}^{2}%
}\frac{d\delta}{\delta}, \label{N1}%
\end{equation}
where $\delta_{e}$ corresponds to $\delta$ calculated at the end of inflation
and it can be obtained by imposing $\epsilon=1$. The expression (\ref{N1}) can
be integrated more easily if we define $x=\delta_{c}/\delta$. In this case%
\begin{equation}
N\approx\frac{3}{4\delta_{c}}%
%TCIMACRO{\dint \limits_{x_{e}}^{x}}%
%BeginExpansion
{\displaystyle\int\limits_{x_{e}}^{x}}
%EndExpansion
\frac{x-2\delta_{c}-2x^{2}\delta_{c}}{\left(  1-x^{2}\right)  x}dx.
\label{N1 x}%
\end{equation}
As we saw previously, physical inflation occurs in the range of%
\[
0<\chi\leq\chi_{c}\Rightarrow\delta_{c}<x\leq1.
\]

Within this range, the integrand of Eq. (\ref{N1 x}) has a divergence at the
point $x=1$ ($\chi=\chi_{c}$). Mathematically, the origin of this divergence
is associated with the existence of the critical point $\left(  \chi,\dot
{\chi}\right)  =\left(  \chi_{c},0\right)  $ in the attractor line of the
phase space\footnote{Note this is a unique feature of the model containing
	$R^{3}$. In fact, for the pure Starobinsky model, this does not happen, since
	$\dot{\chi}$ is always nonzero in the attractor line.}. From a physical point
of view, this divergence must be understood as a limit case that occurs only
when the initial conditions are adjusted with infinite precision so that the
solution passes through the critical point $\left(  \chi,\dot{\chi}\right)
=\left(  \chi_{c},0\right)  $. In this idealized case, upon reaching the
critical point, the scalar field remains in unstable equilibrium and the
universe reaches a pure de Sitter state. Initial conditions adjusted with
infinite precision are not physically achievable. However, by fine-tuning
these conditions, we can increase the number of $e$-folds to an arbitrarily
large value. In Sec. \ref{sec - probabilidade}, we see how accurate this
adjustment must be in order to reach a specific value in the number of $e$-folds.

By integrating Eq. (\ref{N1 x}) in the slow-roll leading-order and considering
$x_{end}\ll1$, we get%
\begin{equation}
N\approx\frac{3}{8\delta_{c}}\ln\left(  \frac{1+x}{1-x}\right)  \approx
\frac{3}{4\delta}\left(  1+\frac{x^{2}}{3}+\frac{x^{4}}{5}+\frac{x^{6}}%
{7}+\cdots\right)  . \label{N}%
\end{equation}
Note that $N$ diverges when $x\rightarrow1$. The expression (\ref{N}) can be
explicitly inverted, resulting in%
\begin{equation}
\delta=\delta_{c}\left[  \frac{\exp\left(  \frac{8\delta_{c}N}{3}\right)
	+1}{\exp\left(  \frac{8\delta_{c}N}{3}\right)  -1}\right]  .
\label{delta N completo}%
\end{equation}

Finally, it is convenient to write the slow-roll parameters $\epsilon$ and
$\eta$ in terms of the number of $e$-folds $N$. By substituting Eq.
(\ref{delta N completo}) in Eqs. (\ref{epsilon 2}) and (\ref{eta 2}), we
obtain%
\begin{equation}
\epsilon\approx\frac{4^{3}}{3}\delta_{c}^{2}\frac{\exp\left(  \frac
	{16\delta_{c}N}{3}\right)  }{\left[  1-\exp\left(  \frac{16\delta_{c}N}%
	{3}\right)  \right]  ^{2}}, \label{epsilon N completo}%
\end{equation}%
\begin{equation}
\eta\approx\frac{16}{3}\delta_{c}\left[  \frac{1+\exp\left(  \frac
	{16\delta_{c}N}{3}\right)  }{1-\exp\left(  \frac{16\delta_{c}N}{3}\right)
}\right]  . \label{eta N completo}%
\end{equation}
It is also worth noting that in a region where $\delta_{c}\ll\delta
\Rightarrow\delta_{c}N\ll1$, the slow-roll parameters
(\ref{epsilon N completo}) and (\ref{eta N completo}) can be approximated by%
\[
\epsilon\approx\frac{3}{4N^{2}}\left[  1-\frac{1}{12}\left(  \frac
{16\delta_{c}N}{3}\right)  ^{2}\right]  ,
\]%
\[
\eta\approx-\frac{2}{N}\left[  1+\frac{1}{12}\left(  \frac{16\delta_{c}N}%
{3}\right)  ^{2}\right]  ,
\]
where $\delta_{c}N$ takes into account the first corrections for Starobinsky inflation.

\subsection{End of inflation and reheating} \label{sec - reaquecimento}

In this subsection, we investigate the end of inflation and the reheating
period. In a relatively general way, this phase can be divided into two parts:
preheating and thermalization. The preheating phase corresponds to the initial
reheating phase, where a large number of matter particles are generated from a
process known as parametric resonance. This process arises assuming that the
transference of energy from the inflaton field to the matter fields occurs
when the inflaton coherently oscillates around the minimum of the potential
\cite{Kofman:1997yn,Bassett:2005xm}. Since preheating is an essentially
nonthermal process, a thermalization stage is necessary for the universe to
reach a radiation era domain with the matter in thermodynamic equilibrium. For
details on the reheating phase, see Refs. \cite{Amin:2014eta,Lozanov:2019jxc}.

The detailed description of the entire reheating period is complex, as it
involves nonperturbative nonequilibrium effects and depends on the
interaction processes between the inflaton and the matter fields. However,
from a phenomenological perspective, we can study the cosmic dynamics of this
period through an equation of state\footnote{In this subsection, we will adopt
	the usual fourth mass dimension for $\rho$. The relation of this quantity with
	$\rho_{\chi}$ is%
	\[
	\rho=\frac{3M_{Pl}^{2}}{2}\rho_{\chi}\text{,}%
	\]
	where%
	\[
	\rho_{\chi}=\frac{1}{2}\dot{\chi}^{2}+V\left(  \chi\right)  .
	\]
	See Eq. (\ref{campo escalar dimensional}) for notation details.}
\begin{equation}
p=w_{re}\left(  N\right)  \rho, \label{EoS}%
\end{equation}
where the number of $e$-folds $N$ takes into account the time dependence of
$w_{re}$. By construction, at the end of reheating, that is, the beginning of
the radiation era, the parameter $w_{re}$ of the equation of state must be
$1/3$. At the beginning of reheating, the parameter $w_{re}$ depends on the
behavior of the potential during the phase of coherent oscillations. Expanding
the potential (\ref{Potencial V(Chi)}) around the minimum $\chi=0$ we have%
\[
V\left(  \chi\right)  \approx\frac{1}{2}V^{\prime\prime}\left(  0\right)
\chi^{2}=\frac{\kappa_{0}}{6}\chi^{2}.
\]

Considering the mean behavior of the scalar field $\chi$ around this minimum
and taking into account that $V\sim\chi^{2}$, we obtain $\left\langle
w_{re}\right\rangle \approx0$ \cite{PhysRevD.28.1243,mukhanov}. Thus, at
the beginning of reheating, the universe behaves as in a matter domination
era. Therefore, during the reheating phase $w_{re}\left(  N\right)  $
transits from $0$ (matter domain) to $1/3$ (radiation era).

An important quantity in the characterization of reheating is the
thermalization temperature $T_{re}$ reached at the beginning of the radiation
era (end of reheating). This temperature is related to the energy density
$\rho_{re}$ through the expression%
\begin{equation}
\rho_{re}=\frac{\pi^{2}}{30}g_{re}T_{re}^{4}\text{,} \label{energy density re}%
\end{equation}
where $g_{re}$ is the effective number of relativistic degrees of freedom at
the end of reheating\footnote{By considering only particles from the standard
	model we have $g_{re}=106.75$ \cite{Husdal:2016haj}.}. On the other hand,
through Eq. (\ref{EoS}) and the covariant conservation equation, we can relate
$\rho_{re}$ with the energy density $\rho_{e}$ at the end of inflation%
\begin{equation}
\dot{\rho}=-3H\rho\left[  1+w_{re}\left(  N\right)  \right]  \Rightarrow
\rho_{re}=\rho_{e}e^{-3N_{re}\left(  1+w_{a}\right)  }\text{,} \label{rho re}%
\end{equation}
with the number of $e$-folds $N_{re}$ characterizing the duration of the
reheating period and
\begin{equation}
w_{a}\equiv\frac{1}{N_{re}}\int_{0}^{N_{re}}w_{re}\left(  N\right)  dN.
\label{w efetivo}%
\end{equation}
Note the quantity $w_{a}$ represents the mean value of the parameter $w_{re}$
during reheating. Also, for $w_{re}$ monotonically
increasing\footnote{Detailed numerical modeling usually indicates a monotonic
	behavior for the reheating equation of state \cite{PhysRevD.97.023533,Maity_2019,PhysRevD.102.103511}.}, we have
\begin{equation}
0\leq w_{a}\leq1/3, \label{limites wa}%
\end{equation}
so the closer to $1/3$, the more effective the rewarming is. By combining Eqs.
(\ref{energy density re}) and (\ref{rho re}) we obtain the reheating
temperature in terms of $w_{a}$ and $N_{re}$%
\begin{equation}
T_{re}=\left(  \frac{30\rho_{e}}{g_{re}\pi^{2}}\right)  ^{\frac{1}{4}%
}e^{-\frac{3}{4}\left(  1+w_{a}\right)  N_{re}}. \label{T re 1}%
\end{equation}

The next step is obtaining $N_{re}$ in terms of cosmological quantities of
interest. For this, it is necessary to "follow" the evolution of the scale $k$
from the moment it crosses the horizon during inflation to the present day.
When crossing the horizon we can write $k=a_{k}H_{k}$ where $a_{k}H_{k}$ is
the comoving Hubble radius at the horizon exit instant. From this relation we
can write \cite{Munoz:2014eqa,Mishra:2021wkm}

\begin{equation}
\frac{k}{a_{0}H_{k}}=\left(  \frac{a_{k}}{a_{e}}\right)  \left(  \frac{a_{e}%
}{a_{re}}\right)  \left(  \frac{a_{re}}{a_{eq}}\right)  \left(  \frac{a_{eq}%
}{a_{0}}\right)  , \label{Escala de Horizonte}%
\end{equation}
where $a_{e}$, $a_{re}$, $a_{eq}$ and $a_{0}$ are the values of the scale
factor at the end of inflation, at the beginning of the radiation era, at the
equivalence era, and in the present day, respectively. Each of the above
fractions represents a specific era of cosmic expansion\footnote{Note that we
	are ignoring the dark energy era domain since it is only relevant very close
	to the present day.}. The fraction $\left(  a_{k}/a_{e}\right)  $ represents
the inflationary period counted from the exit time of the scale $k$ from the
horizon; $\left(  a_{e}/a_{re}\right)  $, the reheating period; $\left(
a_{re}/a_{eq}\right)  $, the radiation era; and $\left(  a_{eq}/a_{0}\right)
$, the matter era until the present day.

The next step is by taking the logarithm of Eq. (\ref{Escala de Horizonte}),
resulting in%
\begin{equation}
\ln\left(  \frac{k}{a_{0}H_{k}}\right)  =-N_{k}-N_{re}+\ln\left(  \frac
{a_{re}}{a_{eq}}\right)  +\ln\left(  \frac{a_{eq}}{a_{0}}\right)
\label{Nre 1}%
\end{equation}
with $N_{re}=\ln\left(  \frac{a_{re}}{a_{e}}\right)  $ and $N_{k}=\ln\left(
\frac{a_{e}}{a_{k}}\right)  $, where this last one is the number of $e$-folds of
the inflationary period measured since the exit of the scale $k$ from the
horizon. The last two terms of Eq. (\ref{Nre 1}) can be rewritten to take into
account the conservation of entropy $S=ga^{3}T^{3}$ in the radiation and
matter eras \cite{Kolb:1990vq,Husdal:2016haj}. In this case, we can
write%
\begin{equation}
T_{eq}=\frac{a_{0}}{a_{eq}}T_{0}\text{ \ and \ }T_{eq}=\left(  \frac{a_{re}%
}{\alpha_{eq}}\right)  \left(  \frac{g_{re}}{g_{0}}\right)  ^{\frac{1}{3}%
}T_{re} \label{Temperaturas}%
\end{equation}
where we consider that the relativistic degrees of freedom in the equivalence
era and in the present day are identical, that is, $g_{0}=g_{eq}$.\footnote{By
	considering only particles from the standard model and disregarding the
	neutrinos masses, we have $g_{0}=3.94$ \cite{Husdal:2016haj}.} By substituting
Eqs. (\ref{T re 1}) and (\ref{Temperaturas}) in Eq. (\ref{Nre 1}), we get%
\begin{align}
N_{re} &  =\frac{4}{3\left(  w_{a}-\frac{1}{3}\right)  }\left\{  N_{k}%
+\ln\left(  \frac{\rho_{e}^{1/4}}{H_{k}}\right)  \right.  +\nonumber\\
&  +\left.  \ln\left[  \left(  \frac{k}{a_{0}T_{0}}\right)  \left(  \frac
{30}{\pi^{2}}\right)  ^{\frac{1}{4}}\left(  \frac{g_{re}}{g_{0}^{4}}\right)
^{\frac{1}{12}}\right]  \right\}  .\label{Nre 2}%
\end{align}

Expressions (\ref{Nre 2}) and (\ref{T re 1}) that determine the duration and
temperature of the reheating are general and hold for a wide range of
inflationary models. By particularizing these expressions for our model, it is
necessary to explicitly determine the energy density at the end of inflation.

In single-field models, $\rho_{e}$ can be written in terms of the potential
$\bar{V}_{e}=\left(  3M_{Pl}^{2}/2\right)  V\left(  \chi_{e}\right)  $ given
in Eq. (\ref{campo escalar dimensional}). By imposing $\epsilon=1$ (end of
inflation) we obtain by Eqs. (\ref{H2 Sta plus R3}), (\ref{Hpto Sta plus R3})
and (\ref{slow roll definitions}) that $V\left(  \chi_{e}\right)  =\dot{\chi
}_{e}^{2}$. Thus,%
\begin{equation}
\rho_{e}=\frac{3}{2}\bar{V}_{e}=\frac{9M_{Pl}^{2}}{4}V\left(  \chi_{e}\right)
\text{.} \label{rho e barra}%
\end{equation}
By Eq. (\ref{Condicao alpha 0 delta c}), we see that $\alpha_{0}$ is a
typically small slow-roll second-order quantity. Thus, at the end of
inflation, terms like $4\alpha_{0}e^{\chi_{e}}$ appearing in the potential are
negligible and Eq. (\ref{Potencial V(Chi)}) can be approximated by the
Starobinsky inflation potential%
\begin{equation}
V\left(  \chi_{e}\right)  \approx\frac{\kappa_{0}}{6}\left(  1-e^{-\chi_{e}%
}\right)  ^{2}. \label{V chi e}%
\end{equation}
An estimate for $\chi_{e}$ is obtained by imposing $\epsilon=1$ on the
approximate slow-roll expression for the parameter $\epsilon$
\begin{equation}
\epsilon\approx\frac{1}{3}\left(  \frac{V^{\prime}\left(  \chi_{e}\right)
}{V\left(  \chi_{e}\right)  }\right)  ^{2}=1\Rightarrow e^{-\chi_{e}}%
=2\sqrt{3}-3. \label{chi e}%
\end{equation}
Thus, substituting Eqs. (\ref{chi e}) and (\ref{V chi e}) into Eq.
(\ref{rho e barra}), we obtain the energy density at the end of inflation
\begin{equation}
\rho_{e}\approx\frac{3\zeta^{4}}{4}\left(  2-\sqrt{3}\right)  ^{2}M_{Pl}^{4},
\label{rho e barra 1}%
\end{equation}
where the dimensionless parameter $\zeta$ is defined as%
\begin{equation}
\zeta\equiv\left(  \frac{\kappa_{0}}{M_{Pl}^{2}}\right)  ^{\frac{1}{4}}.
\label{xi}%
\end{equation}

Therefore, considering $H_{k}$ in the slow-roll leading-order ($H_{k}%
^{2}\approx\kappa_{0}/12$) and substituting Eq. (\ref{rho e barra 1}) in Eqs.
(\ref{Nre 2}) and (\ref{T re 1}) we get
\begin{equation}
N_{re}=\frac{4}{3\left(  w_{a}-\frac{1}{3}\right)  }\left[  N_{k}%
+0.79+\ln\left(  \frac{k}{a_{0}T_{0}\zeta}\right)  +\frac{1}{12}\ln\left(
\frac{g_{re}}{g_{0}^{4}}\right)  \right]  , \label{Nre 3}%
\end{equation}
and%
\begin{equation}
T_{re}=\frac{0.64\zeta}{g_{re}^{1/4}}e^{-\frac{3}{4}\left(  1+w_{a}\right)
	N_{re}}M_{Pl}. \label{T re 2}%
\end{equation}
We will see in the next section how constraints on $N_{re}$ and $T_{re}$ restrict
the range of the number of $e$-folds $N_{k}$ of inflation.

\section{Observational constraints} \label{sec - obs}

Based on the developments of the previous sections, we now establish constraints to
the proposed inflationary model. These constraints follow different approaches and
are related to periods before, during, and after the inflationary regime.

\subsection{Range in the number of $e$-folds $N_{k}$ of inflation}

The description of the reheating period as presented in Sec.
\ref{sec - reaquecimento} with the assumption of later eras of radiation and
matter dominance allows us to establish an interval for the number of $e$-folds
$N_{k}$ of inflation. For this, it is necessary to provide some details about
the minimum characteristics of the reheating period to be considered.

The cosmological period comprising energy scales above $10^{4}$ $%
%TCIMACRO{\unit{GeV}}%
%BeginExpansion
\operatorname{GeV}%
%EndExpansion
$ until the end of inflation $10^{15}-10^{16}$ $%
%TCIMACRO{\unit{GeV}}%
%BeginExpansion
\operatorname{GeV}%
%EndExpansion
$ is an uncertain period for particle physics. In fact, extensions of the
standard model together with possible nonminimal couplings can provide extra
processes of coupling the inflaton with the matter fields. However, even if
these new processes are not present (or are not relevant for reheating) we can
adopt as a minimum hypothesis the usual couplings of the standard matter fields
with gravity. In this context, already in the Einstein frame, this means
that the inflaton field $\phi$ decays predominantly in the Higgs doublet $h$,
and in next-to-leading order, in a pair of gluons. By considering these two main decay
channels, the inflaton decay rate is given by \cite{Gorbunov:2012ns,Bernal:2020qyu}
\begin{equation}
\Gamma_{\phi}\approx\frac{1}{24\pi}\left[  \left(  1-6\xi\right)  ^{2}%
+\frac{49\alpha_{s}^{2}}{4\pi^{2}}\right]  \frac{m_{\phi}^{3}}{M_{Pl}^{2}},
\label{Gamma phi}%
\end{equation}
where $\xi$ is the nonminimum coupling constant between the Higgs field and
gravitation, that is $\xi\left\vert h\right\vert ^{2}R$,\footnote{The
	nonminimum coupling is necessary due to issues of the renormalizability of
	scalar fields in curved spaces \cite{CALLAN197042}.} $\alpha_{s}$ is the
coupling constant of QCD on the reheating energy scale\footnote{By
	extrapolating the experimental results of the running coupling of $\alpha_{s}$
	\cite{Proceedings:2015eho}, we can estimate that during the reheating period
	$0.01\lesssim\alpha_{s}\lesssim0.03$.} and $m_{\phi}$ is the effective mass of
the inflaton, given by%
\begin{equation}
m_{\phi}^{2}\equiv V^{\prime\prime}\left(  0\right)  =\frac{\zeta^{4}}%
{3}M_{Pl}^{2}. \label{m phi}%
\end{equation}

The next step is determining $\zeta$ in terms of the model parameters and the
scalar amplitude $A_{s}$ measured by the Planck satellite
\cite{Planck:2018vyg}. Considering the pivot scale $k=0.002$ $Mpc^{-1}$ as the scale of interest \footnote{We use this
			pivot scale in agreement with the observations presented in Refs.
			\cite{Akrami:2018odb,BICEPKeck:2021gln} associated with observations
		$n_{s}\times r_{0.002}$.}, we have in slow-roll
leading-order \cite{Baumann:2018muz}%
\begin{equation}
A_{s}=\frac{3}{16\pi^{2}M_{p}^{2}}\frac{V_{k}^{3}}{V_{k}^{\prime2}},
\label{As}%
\end{equation}
where the scalar amplitude $A_{s}^{0.002}=2.3\times10^{-9}$ is
obtained from the expression $A_{s}^{k}=A_{s}^{\ast}\left(  k/k^{\ast
}\right)  ^{n_{s}-1}$ taking into account that $A_{s}^{\ast}%
=2.1\times10^{-9}$, $k^{\ast}=0.05$ $Mpc^{-1}$ and %
$n_{s}=0.9665$. See Ref. \cite{Planck:2018vyg} for details.

Thus, substituting Eqs. (\ref{V aprox}) and
(\ref{V linha aprox}) into Eq. (\ref{As}), we obtain%
\[
\kappa_{0}\approx2^{7}\pi^{2}M_{Pl}^{2}A_{s}^{0.002}\left(  \delta_{k}%
-\frac{1}{3}\delta_{k}^{-1}\alpha_{0}\right)  ^{2}.
\]
Furthermore, we can rewrite the last equation in terms of the number of
$e$-folds $N_{k}$ and the parameter $\zeta$ given in Eqs.
(\ref{delta N completo}) and (\ref{xi}), respectively%
\begin{equation}
\zeta^{4}\approx2^{11}\pi^{2}A_{s}^{0.002}\left[  \frac{\delta_{c}\exp\left(
	\frac{8}{3}\delta_{c}N_{k}\right)  }{\exp\left(  \frac{16}{3}\delta_{c}%
	N_{k}\right)  -1}\right]  ^{2},\label{zeta 4}%
\end{equation}
remembering that $\alpha_{0}\approx3\delta_{c}^{2}$. In the case of
$\delta_{c}\rightarrow0$, we recover the well-known relation of the
Starobinsky model $\zeta^{4}\approx72\pi^{2}A_{s}^{0.002}N_{k}^{-2}$.

By having determined the decay rate of the inflaton, we can estimate the
reheating temperature $T_{re}$ as follows: the thermal equilibrium of the
cosmic fluid is reached when $\Gamma_{\phi}\sim H$ \cite{Kofman:1997yn}. Thus, the reheating
temperature can be obtained from $3M_{Pl}^{2}H_{re}^{2}=\rho_{re}=\pi
^{2}g_{re}T_{re}^{4}/30$, which results in%
\begin{equation}
T_{re}\approx0.5\sqrt{\Gamma_{\phi}M_{Pl}}\approx6\times10^{16}\zeta^{3}%
\sqrt{\left(  1-6\xi\right)  ^{2}+\frac{49\alpha_{s}^{2}}{4\pi^{2}}}\text{ }%
%TCIMACRO{\unit{GeV}}%
%BeginExpansion
\operatorname{GeV}%
%EndExpansion
, \label{T re 3}%
\end{equation}
where we use $g_{re}=106.75$ and $M_{Pl}=2.4\times10^{18}%
%TCIMACRO{\unit{GeV}}%
%BeginExpansion
\operatorname{GeV}%
%EndExpansion
$. Note the value of $T_{re}$ above depends on the inflationary parameters
through $\zeta$. Substituting Eq. (\ref{zeta 4}) into Eq. (\ref{T re 3}), we
obtain%
\begin{align}
T_{re}  &  \approx3.4\times10^{13}\left[  \frac{\delta_{c}\exp\left(  \frac
	{8}{3}\delta_{c}N_{k}\right)  }{\exp\left(  \frac{16}{3}\delta_{c}%
	N_{k}\right)  -1}\right]  ^{\frac{3}{2}}\times\nonumber\\
&  \times\sqrt{\left(  1-6\xi\right)  ^{2}+\frac{49\alpha_{s}^{2}}{4\pi^{2}}%
}\text{ }%
%TCIMACRO{\unit{GeV}}%
%BeginExpansion
\operatorname{GeV}%
%EndExpansion
. \label{T re 4}%
\end{align}
Based on Eq. (\ref{T re 4}), we can estimate the order of magnitude of the
reheating temperature. Considering $N_{k}=50$ and $\delta_{c}=0$
(Starobinsky), we have \cite{Gorbunov:2012ns}%
\[
T_{re}\sim\left\{
\begin{array}
[c]{c}%
10^{10}\left\vert 1-6\xi\right\vert \text{ }%
%TCIMACRO{\unit{GeV}}%
%BeginExpansion
\operatorname{GeV}%
%EndExpansion
\text{\ \ if \ }\xi\neq1/6\\
10^{8}\text{ }%
%TCIMACRO{\unit{GeV}}%
%BeginExpansion
\operatorname{GeV}%
%EndExpansion
\text{\ \ if \ }\xi=1/6
\end{array}
\right.  ,
\]
where we adopt $\alpha_{s}\sim10^{-2}$. Although $T_{re}$ depends on the value
of $\xi$,\footnote{The debate over the range of allowable values for $\xi$ is
	complex and is directly related to the stability of the Higgs field at high
	energies. This analysis depends on several factors such as the mass value of
	the top quark and the running coupling of $\xi$ in a Friedmann background
	\cite{Herranen:2015ima,Markkanen:2018pdo}.} the expression above gives a
minimum estimate for $T_{re}$. In fact, even though the Higgs field has a
conformal coupling with gravitation, that is $\xi=1/6$, the inflaton decay
channel in two gluons gives $T_{re}\sim10^{8}$ $%
%TCIMACRO{\unit{GeV}}%
%BeginExpansion
\operatorname{GeV}%
%EndExpansion
$.

The previous construction allows establishing a minimum temperature
$T_{re}^{\left(  \min\right)  }$ relatively independent of the details of the
reheating phase. Even considering minimal hypotheses for reheating and a
conformal coupling of the Higgs field with gravitation, we have%
\begin{align}
T_{re}^{\left(  \min\right)  } &  \approx3.4\times10^{13}\left(  \frac
{7\alpha_{s}}{2\pi}\right)  \left[  \frac{\delta_{c}\exp\left(  \frac{8}%
	{3}\delta_{c}N_{k}\right)  }{\exp\left(  \frac{16}{3}\delta_{c}N_{k}\right)
	-1}\right]  ^{\frac{3}{2}}\text{ }%
%TCIMACRO{\unit{GeV}}%
%BeginExpansion
\operatorname{GeV}%
%EndExpansion
\nonumber\\
&  \approx3.8\times10^{11}\left[  \frac{\delta_{c}\exp\left(  \frac{8}%
	{3}\delta_{c}N_{k}\right)  }{\exp\left(  \frac{16}{3}\delta_{c}N_{k}\right)
	-1}\right]  ^{\frac{3}{2}}\text{ }%
%TCIMACRO{\unit{GeV}}%
%BeginExpansion
\operatorname{GeV}%
%EndExpansion
.\label{T re min}%
\end{align}

Since we characterized the reheating phase, let us see how the expressions of
$N_{re}$ and $T_{re}$, obtained at the end of Sec.
\ref{sec - reaquecimento} provide restrictions for the number of $e$-folds
$N_{k}$.

By considering the scale of interest $k=0.002$ $Mpc^{-1}$, $T_{0}=2.73$ $%
%TCIMACRO{\unit{K}}%
%BeginExpansion
\operatorname{K}%
%EndExpansion
$, $g_{re}=106.75$ and $g_{0}=3.94$, we can rewrite Eq. (\ref{Nre 3})
as\footnote{Recovering the units, we have $\log\left(  \frac{k}{a_{0}T_{0}%
	}\right)  =\ln\left(  \frac{kc\hbar}{a_{0}T_{0}k_{B}}\right)  =-65.079$.}%
\begin{align}
N_{re} &  =\frac{4}{1-3w_{a}}\left[  64.36-N_{k}+\ln\left(  \zeta\right)
\right]  \nonumber\\
&  =\frac{4}{1-3w_{a}}\left\{  61.87-N_{k}+\frac{1}{2}\ln\left[  \frac
{\delta_{c}\exp\left(  \frac{8}{3}\delta_{c}N_{k}\right)  }{\exp\left(
	\frac{16}{3}\delta_{c}N_{k}\right)  -1}\right]  \right\}  .\label{Nre vinculo}%
\end{align}
The upper limit for $N_{k}$ is obtained taking into account that for physical
consistency $N_{re}\geq0$. So, due to Eq. (\ref{limites wa}), we have%
\begin{equation}
N_{k}-\frac{1}{2}\ln\left[  \frac{\delta_{c}\exp\left(  \frac{8}{3}\delta
	_{c}N_{k}\right)  }{\exp\left(  \frac{16}{3}\delta_{c}N_{k}\right)
	-1}\right]  \leq61.87. \label{inequacao Nre}%
\end{equation}
Table \ref{tab:upper-limits} shows the upper limits of $N_{k}$ for different values of $\alpha_{0}%
$.%
\begin{table}[h]
	\caption{\label{tab:upper-limits} Maximum values for $N_{k}$ considering different values of
		$\alpha_{0}$. Note that the existence of a minimum slow-roll period consistent
		with the approximations made in Sec. \ref{sec - slow roll} requires
		$\alpha_{0}\lesssim10^{-3}$.}
\begin{tabular}
[c]{|c|c|c|}\hline 
$\alpha_{0}$ & $\delta_{c}$ & $N_{k}$\\\hline \hline
$0$ & $0$ & $58.99$\\\hline
$10^{-5}$ & $1.85\times10^{-3}$ & $58.99$\\\hline
$10^{-4}$ & $5.8\times10^{-3}$ & $58.92$\\\hline
$5\times10^{-4}$ & $1.3\times10^{-2}$ & $58.69$\\\hline
$10^{-3}$ & $1.83\times10^{-2}$ & $58.45$\\\hline
\end{tabular}
\end{table}

In turn, the lower limit for the number of $e$-folds $N_{k}$ is obtained taking
into account that $T_{re}\gtrsim T_{re}^{\left(  \min\right)  }$. Thus, by
using Eqs. (\ref{T re 2}), (\ref{zeta 4}), (\ref{T re min}) and
(\ref{Nre vinculo}), we have%
\begin{equation}
\left[  \frac{\exp\left(  \frac{16}{3}\delta_{c}N_{k}\right)  -1}{\delta
	_{c}\exp\left(  \frac{8}{3}\delta_{c}N_{k}\right)  }\right]  \exp\left[
-\frac{3\left(  1+w_{a}\right)  }{1-3w_{a}}\bar{N}_{re}\right]  \gtrsim
10^{-5}, \label{inequacao Tre}%
\end{equation}
where%
\begin{equation}
\bar{N}_{re}=61.87-N_{k}+\frac{1}{2}\ln\left[  \frac{\delta_{c}\exp\left(
	\frac{8}{3}\delta_{c}N_{k}\right)  }{\exp\left(  \frac{16}{3}\delta_{c}%
	N_{k}\right)  -1}\right]  . \label{Nre barra}%
\end{equation}
By expression (\ref{Nre barra}), we see that as $N_{k}$ decreases, $\bar
{N}_{re}$ increases and consequently the exponential that contains $\bar
{N}_{re}$ reduces the value of the left-hand side (lhs) of the expression (\ref{inequacao Tre}).
The decay speed of this exponential depends on the coefficient
\[
3<\frac{3\left(  1+w_{a}\right)  }{1-3w_{a}}<\infty,
\]
where the lower and upper limits were established from Eq. (\ref{limites wa}).
Thus, for the lower limit in $N_{k}$ to be robust, i.e., relatively
independent of the reheating period details, we must adopt $w_{a}\rightarrow0$
in Eq. (\ref{inequacao Tre}).\footnote{Note that when $w_{a}\rightarrow0$, the
	thermalization process proceeds as slowly as possible.} In this case, the
expression (\ref{inequacao Tre}) is rewritten as%

\begin{widetext}
\begin{equation}
\left[  \frac{\exp\left(  \frac{16}{3}\delta_{c}N_{k}\right)  -1}{\delta
	_{c}\exp\left(  \frac{8}{3}\delta_{c}N_{k}\right)  }\right]  \exp\left\{
-3\left[  61.87-N_{k}+\frac{1}{2}\ln\left(  \frac{\delta_{c}\exp\left(
	\frac{8}{3}\delta_{c}N_{k}\right)  }{\exp\left(  \frac{16}{3}\delta_{c}%
	N_{k}\right)  -1}\right)  \right]  \right\}  \gtrsim10^{-5}.
\label{inequacao Tre 1}%
\end{equation}
\end{widetext}

Table \ref{tab:lower-limits} shows lower limits of $N_{k}$ for different values of $\alpha_{0}$.%

\begin{table}[h]
	\caption{\label{tab:lower-limits} Minimum values for $N_{k}$ considering different values of
		$\alpha_{0}$. As in the previous case, $\alpha_{0}$ is limited by $\alpha
		_{0}\lesssim10^{-3}$.}
\begin{tabular}
[c]{|c|c|c|}\hline
$\alpha_{0}$ & $\delta_{c}$ & $N_{k}$\\\hline \hline
$0$ & $0$ & $53.32$\\\hline
$10^{-5}$ & $1.85\times10^{-3}$ & $53.31$\\\hline
$10^{-4}$ & $5.8\times10^{-3}$ & $53.23$\\\hline
$5\times10^{-4}$ & $1.3\times10^{-2}$ & $52.91$\\\hline
$10^{-3}$ & $1.83\times10^{-2}$ & $52.57$\\\hline
\end{tabular}
\end{table}

We see from Tables \ref{tab:upper-limits} and \ref{tab:lower-limits} that for $\alpha_{0}\leq10^{-3}$, the variation
of $N_{k}$ is at most $0.75$. Therefore, based on these tables and considering
the maximum variation allowed for $N_{k}$, we adopt the following range for
the number of $e$-folds:%
\begin{equation}
52\leq N_{k}\leq59\text{.} \label{Intervalo Nk}%
\end{equation}
Although it is not a very restrictive range, it was obtained from very general
considerations. In this sense, the above result is quite robust and little
dependent on the details of the reheating, radiation and matter eras that
characterize the postinflationary universe. This range of $N_{k}$ will be
used in the next sections.

\bigskip

The previous modeling was developed using the thermal equilibrium
temperature. However, in some inflationary models, the reheating temperature 
$T_{re}$ is established long after the equilibrium between the decay rate $%
\Gamma _{\phi }$ and the expansion $H$ \cite{Allahverdi:2005fq,Allahverdi:2005mz,Allahverdi:2010xz}. In this case, we can still obtain a
result identical to Eq. (\ref{Intervalo Nk}) but using a different approach. We can treat the
problem not through a reheating temperature $T_{re}$, given in Eq.
(\ref{T re 3}), but through a quasithermal energy density and decay rate of a universe in a quasithermal phase. In this case, the relation $\rho
_{qt}=3M_{Pl}^{2}\Gamma_{\phi}^{2}$ is preserved, and similarly to what was
developed previously, we can establish a minimum quasithermal energy
density%
\begin{equation}
\rho_{qt}^{\left(  \min\right)  }\approx10^{48}\left[  \frac{\delta_{c}%
	\exp\left(  \frac{8}{3}\delta_{c}N_{k}\right)  }{\exp\left(  \frac{16}%
	{3}\delta_{c}N_{k}\right)  -1}\right]  ^{6}%
%TCIMACRO{\unit{GeV}}%
%BeginExpansion
\operatorname{GeV}%
%EndExpansion
^{4},
\end{equation}
wherewith we establish the robust result for the lower limit of the number of
$e$-folds $N_{k}$ and, consequently, the restriction Eq. (\ref{Intervalo Nk}).

\subsection{Constraints obtained from CMB anisotropies} \label{sec - CMB}

One of the main motivations for the inflationary paradigm is the generation of
causally connected initial fluctuations that provide the initial conditions
for the structure formation process. These fluctuations arise from the
quantization of the metric and inflaton field(s) perturbations
\cite{MUKHANOV1992203,mukhanov}.

In most cases, the relevant perturbations produced by inflationary models are
the scalar and tensor ones. While the scalar perturbations are responsible for
generating the primordial inhomogeneities, the tensor ones (gravitational
waves) carry information directly from the inflationary period. In the
simplest models involving only a canonical scalar field minimally coupled with
general relativity\footnote{This is exactly our case when described in
	Einstein's frame.}, the scalar and tensor perturbations are dominantly
characterized by four parameters: scalar and tensor amplitudes $A_{s}$ and
$A_{t}$, respectively; and scalar and tensor spectral indices, $n_{s}$ and
$n_{t}$ respectively. Of these four parameters, only three are independent,
since $n_{t}$ can be written as a combination of $A_{s}$ and $A_{t}$.

From the point of view of current observations, the anisotropies in the CMB
measure scalar quantities and establish an upper limit for tensor quantities
\cite{Akrami:2018odb,BICEPKeck:2021gln}. In this context, it is
convenient to describe the upper constraint with the tensor-scalar ratio
$r\equiv A_{t}/A_{s}$.

In the slow-roll leading-order, we have \cite{Longden:2017iei,Baumann:2018muz}
\begin{equation}
n_{s}=1+\eta-2\epsilon\text{ \ and \ }r=16\epsilon, \label{ns e r}%
\end{equation}
where $\epsilon$ and $\eta$ are the slow-roll parameters\footnote{The scalar
	amplitude $A_{s}$ is given by Eq. (\ref{As}).}. For the proposed model,
$\epsilon$ and $\eta$ are given in Eqs. (\ref{epsilon N completo}) and
(\ref{eta N completo}) and therefore they depend on the number of $e$-folds
$N_{k}$ and the parameter $\alpha_{0}\approx3\delta_{c}^{2}$.

\begin{figure}[h] 
	\centering
	\includegraphics[scale=0.37]{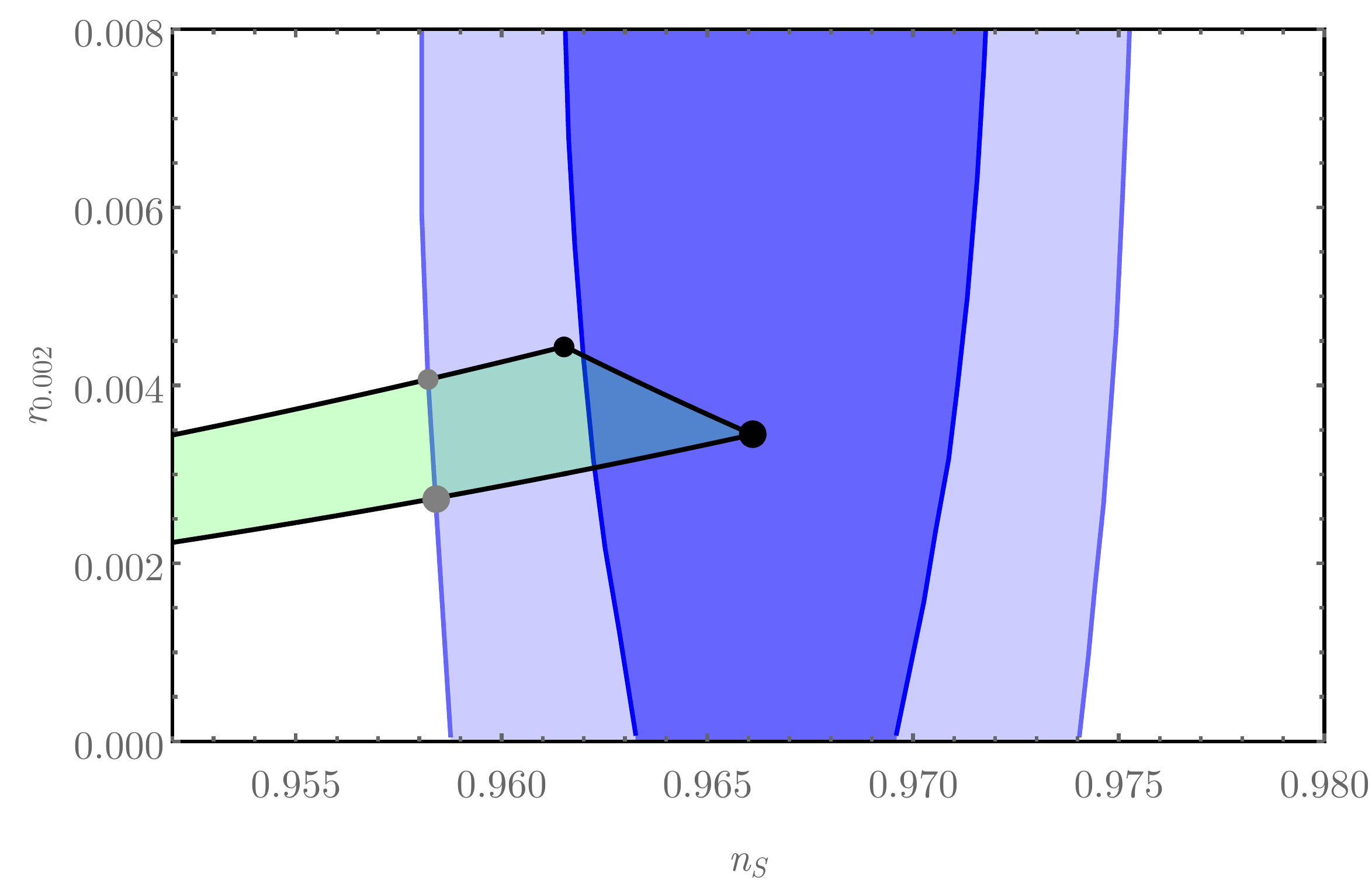}
	\caption{\label{fig:plano-nsxr} The blue contours correspond to $68\%$ and $95\%$ C.L.
		constraints on $n_{s}\times r_{0.002}$ given by Planck plus BICEP3/Keck plus BAO data
		\cite{BICEPKeck:2021gln}. The black circles represent Starobinsky model
		($\alpha_{0}=0$) for $N_{k}=52$ (smaller one) and $N_{k}=59$ (bigger one). As
		$\alpha_{0}$ increases the curves move to the left (light green region)
		decreasing the tensor-to-scalar ratio and the scalar tilt values. The grey
		circles represent the upper limits for $\alpha_{0}$ associated with $95\%$
		C.L.. In this case, $N_{k}=52$ corresponds to $\alpha_{0}=4.1\times10^{-5}$
		and $r=4.1\times10^{-3}$; and $N_{k}=59$ corresponds to $\alpha_{0}%
		=8.7\times10^{-5}$ and $r=2.7\times10^{-3}$.}
\end{figure}

Figure \ref{fig:plano-nsxr} shows the parameter space $n_{s}\times r_{0.002}$ containing the
observational constraints (blue) obtained from Ref. \cite{BICEPKeck:2021gln}
and the theoretical evolution of the model (light green) obtained from Eq.
(\ref{ns e r}). The analysis is done considering the interval in Eq.
(\ref{Intervalo Nk}) for the number of $e$-folds.%

The black dots in Fig. \ref{fig:plano-nsxr} represent the Starobisnky solution, where
$\alpha_{0}=0$. As $\alpha_{0}$ increases, the region predicted by the model
(light green) shifts to the left and slightly downwards. The $95\%$ C.L. is
represented by the straight line joining the gray dots. This boundary
establishes maximum values allowed for $\alpha_{0}$ being the largest of them
$8.7\times10^{-5}$. Thus, we can consider that the CMB
observations upper limit the value of $\alpha_{0}$ to $10^{-4}$.

Similar results were obtained in Refs. \cite{Huang:2013hsb,Cheong:2020rao} by considering the $R^{3}$ term as a small
correction to Starobinsky inflation.

\subsection{Initial conditions} \label{sec - probabilidade}

As shown in Sec. \ref{sec - pot}, the existence of a maximum point in the
potential limits the initial conditions that drive inflation. In this sense,
the purpose of this subsection is to estimate how severe this limitation is.

By studying the phase space, exemplified in Fig. \ref{fig:phase-space}, we see that any set of
initial conditions $\left(  \chi_{ini},\dot{\chi}_{ini}\right)  $ that reaches
the attractor line to the right of the critical point $\left(  \chi
_{c},0\right)  $ leads to a regime of accelerated expansion that never ends.
Furthermore, we also note that the future dynamics of the scalar field $\chi$
depends essentially on the initial condition $\chi_{ini}$. In fact, the
near-vertical direction fields in Fig. \ref{fig:phase-space} imply that the value of
$\dot{\chi}_{ini}$ has little influence on the subsequent dynamics of the
field $\chi$. Based on this observation, we can assume (approximately) that
only an initial condition $\chi_{ini}<\chi_{c}$ leads to a physical inflation,
that is, inflation with a graceful exit.

The limitation on $\chi_{ini}$ leads to a limitation on the curvature scalar
$R_{ini}$. In fact, by Eq. (\ref{Chi funcao R}), we can write%
\begin{equation}
R\left(  \chi\right)  =\zeta^{4}\left(  \frac{-1+\sqrt{1+4\alpha_{0}\left(
		e^{\chi}-1\right)  }}{2\alpha_{0}}\right)  M_{Pl}^{2}, \label{R chi}%
\end{equation}
and so, $\chi=\chi_{c}$ results in a maximum value for the curvature scalar
$R_{\max}=R\left(  \chi_{c}\right)  $ still consistent with a physical
inflation. Therefore, whatever the preinflationary cosmological model, it
must provide as an initial condition $R_{ini}<R_{\max}$.

Let us see next how $R_{\max}$ depends on $\alpha_{0}$ and $N_{k}$.
Substituting Eqs. (\ref{Chi c}) and (\ref{zeta 4}) into Eq. (\ref{R chi}), we
obtain%
\begin{align*}
R_{\max}\left(  \alpha_{0},N_{k}\right)   &  \simeq1.5\times10^{-5}\alpha
_{0}\left[  \frac{\exp\left(  \frac{8}{3}\sqrt{\frac{\alpha_{0}}{3}}%
	N_{k}\right)  }{\exp\left(  \frac{16}{3}\sqrt{\frac{\alpha_{0}}{3}}%
	N_{k}\right)  -1}\right]  ^{2}\times\\
&  \times\left(  \frac{-1+\sqrt{1+12\alpha_{0}+4\sqrt{3\alpha_{0}}}}%
{2\alpha_{0}}\right)  M_{Pl}^{2}.
\end{align*}
As $\alpha_{0}$ decreases, $R_{\max}$ increases and at the limit $\alpha
_{0}\rightarrow0$ (Starobinsky), we obtain $R_{\max}\rightarrow\infty$. In the
following table, we show $R_{\max}$ for different values of $\alpha_{0}$
considering $N_{k}=52$ and $N_{k}=59$:%
\begin{table}[h]
	\caption{\label{tab:values-a-R} Values of $R_{\max}$ varying $\alpha_{0}$ to $N_{k}=52$ and
		$N_{k}=59$. Due to the results of Sec. \ref{sec - CMB}, we consider
		$\alpha_{0}\leq10^{-4}$. Note that for $\alpha_{0}\sim10^{-18}$, the quantity
		$R_{\max}$ reaches the Planck scale.}
\begin{tabular}
[c]{|c|c|c|}\hline
$\alpha_{0}$ & $R_{\max}^{52}\left(  M_{Pl}^{2}\right)  $ & $R_{\max}%
^{59}\left(  M_{Pl}^{2}\right)  $\\\hline  \hline
$0$ & $\infty$ & $\infty$\\\hline
$10^{-18}$ & $1.01$ & $0.79$\\\hline
$10^{-5}$ & $3\times10^{-7}$ & $2\times10^{-7}$\\\hline
$10^{-4}$ & $8\times10^{-8}$ & $6\times10^{-8}$\\\hline
\end{tabular}
\end{table}

The probabilistic analysis that determines whether a generic $R_{ini}$
produces physical inflation is not trivial. In fact, this analysis depends
on the probability distribution of the curvature scalar in preinflationary
models. However, if a uniform distribution is assumed for $R_{ini}$, the
results in Table \ref{tab:values-a-R} indicate severe limitations for the allowed initial
conditions. For example, for $\alpha_{0}\sim10^{-5}$, we must have
$R_{ini}\lesssim10^{-7}M_{Pl}^{2}$, which corresponds to a probability of
$10^{-5}$ $\%$ that a generic $R_{ini}$ leads to a physical inflationary
regime\footnote{For this estimate, we consider that $R_{ini}/M_{Pl}^{2}$ has
	an equal probability of assuming a value between $0$ and $1$. The upper limit
	of $1$ takes into account the ignorance of the theory of gravitation on the
	Planck scale.}. This probability increases as $\alpha_{0}$ decreases. However,
for very small $\alpha_{0}$, the $R^{3}$ term in the action becomes negligible
and the model essentially behaves like the Starobinsky one. Note that this
kind of limitation can be avoided if the pre-inflationary model has some
dynamic mechanism that delays the start of inflation such that $R_{ini}$ is
always smaller than $R_{\max}$.

In addition to an upper limit, the initial condition $R_{ini}$ also has a
lower limit related to the need for a sufficiently long inflationary period $N$.
In fact, due to the constraint (\ref{Intervalo Nk}), the inflationary regime
must satisfy the condition $N>N_{k}$. By understanding the restrictions
imposed by this condition, let us consider the region of the attractor line
where physical inflation occurs. We know that the total "length" of this
region is $\chi_{c}-\chi_{e}$ where $\chi_{e}$ is given by Eq. (\ref{chi e}).
On the other hand, we can select a subregion in which the number of $e$-folds
is greater than or equal to a fixed $N_{k}$. The length of this subregion is
given by $\chi_{c}-\chi_{k}$. So we define a probability function%
\begin{equation}
P\left(  \alpha_{0},N_{k}\right)  =\frac{\chi_{c}-\chi_{k}}{\chi_{c}-\chi_{e}%
}, \label{Prob}%
\end{equation}
which (approximately) measures how likely arbitrary initial conditions lead to
an inflation with a number of $e$-folds $N$ greater than or equal to $N_{k}$.
Note that in this analysis, we are excluding the entire set of initial
conditions that lead $\chi$ to the attractor line to the right side of the
critical point $\left(  \chi_{c}\text{,}0\right)  $. See Fig. \ref{fig:phase-space} for details.

Substituting Eqs. (\ref{Chi c}), (\ref{delta N completo}) and (\ref{chi e})
into Eq. (\ref{Prob}), we get%
\begin{equation}
P\left(  \alpha_{0},N_{k}\right)  \approx\frac{\ln\left[  \frac{\exp\left(
		\frac{8N_{k}}{3}\sqrt{\frac{\alpha_{0}}{3}}\right)  +1}{\exp\left(
		\frac{8N_{k}}{3}\sqrt{\frac{\alpha_{0}}{3}}\right)  -1}\right]  }{\ln\left(
	\frac{6-3\sqrt{3}}{\sqrt{\alpha_{0}}}\right)  }.
\label{Probabilidade completa}%
\end{equation}
In Table \ref{tab:probability}, we calculate $P\left(  \alpha_{0},N_{k}\right)  $ for different
values of $\alpha_{0}$ considering $N_{k}=52$ and $N_{k}=59$:%
\begin{table}[h]
	\caption{\label{tab:probability} Probability $P\left(  \alpha_{0},N_{k}\right)  $ for different
		values of $\alpha_{0}$ with $N_{k}=52$ and $N_{k}=59$. Again, we consider
		$\alpha_{0}\leq10^{-4}$.}
\begin{tabular}
[c]{|c|c|c|}\hline
$\alpha_{0}$ & $P\left(  \alpha_{0},52\right)  $ & $P\left(  \alpha
_{0},59\right)  $\\\hline \hline
$0$ & $\infty$ & $\infty$\\\hline
$10^{-18}$ & $0.83$ & $0.82$\\\hline
$10^{-5}$ & $0.37$ & $0.35$\\\hline
$10^{-4}$ & $0.22$ & $0.19$\\\hline
\end{tabular}
\end{table}

The results in Table \ref{tab:probability} show that the probability of obtaining an
inflationary period with at least $N_{k}$ $e$-folds increases as $\alpha_{0}$
decreases. Even so, for $\alpha_{0}\sim10^{-5}$ this probability does not
reach $40\%$.

Still related to the initial conditions, a third point to be highlighted
concerns the inflationary regime in the vicinity of the critical point
$\chi_{c}$. In that neighborhood, quantum fluctuations dominate over the
classical slow-roll regime, and the universe reaches an eternal inflation
regime \cite{PhysRevD.27.2848,Linde:1986fd}.

According to Ref. \cite{Creminelli:2008es} quantum fluctuations dominate whenever%
\begin{equation}
\pi^{2}M_{Pl}^{2}\frac{\dot{\chi}^{2}}{H^{4}}<1.\label{condA}%
\end{equation}
In the slow-roll leading-order regime, we can rewrite the above condition as \cite{book:145101}
\begin{equation}
\frac{9}{8\pi^{2}M_{Pl}^{2}}\frac{V^{3}}{V^{\prime2}}>1.\label{condB}%
\end{equation}
Substituting Eqs. (\ref{V aprox}), (\ref{V linha aprox}) and (\ref{zeta 4}) in
Eq. (\ref{condB}), we get%
\begin{equation}
\left\vert \delta-\frac{\alpha_{0}}{3}\delta^{-1}\right\vert <Q\sqrt
{\frac{\alpha_{0}}{3}},\label{condC}%
\end{equation}
where%

\[
Q\equiv\sqrt{24A_{s}^{0.002}}\left[  \frac{\exp\left(  \frac{8}{3}\sqrt
	{\frac{\alpha_{0}}{3}}N_{k}\right)  }{\exp\left(  \frac{16}{3}\sqrt
	{\frac{\alpha_{0}}{3}}N_{k}\right)  -1}\right]  .
\]
By analyzing the condition (\ref{condC}), we conclude that the eternal
inflation regime is reached whenever $\delta=e^{-\chi}$ is within the interval%
\begin{equation}
\delta^{\left(  2\right)  }<\delta<\delta^{\left(  1\right)  },\label{condD}%
\end{equation}
where%
\begin{align*}
\delta^{\left(  1\right)  } &  =\left(  Q+\sqrt{Q^{2}+1}\right)  \sqrt
{\frac{\alpha_{0}}{3}},\\
\delta^{\left(  2\right)  } &  =\left(  -Q+\sqrt{Q^{2}+1}\right)  \sqrt
{\frac{\alpha_{0}}{3}}.
\end{align*}

Considering the intervals $52\leq N_{k}\leq59$ and $\alpha_{0}\leq10^{-4}$, a
numerical analysis shows that the values of $\delta^{\left(  1\right)  }$ and
$\delta^{\left(  2\right)  }$ depend only on the value of $\alpha_{0}$. In
Table \ref{tab:delta-et-inflation}, we calculate $\delta^{\left(  1\right)  }$ and $\delta^{\left(
	2\right)  }$ for different values of $\alpha_{0}$.

\begin{table}[h]
	\caption{\label{tab:delta-et-inflation} Calculation of $\delta^{\left(  1\right)  }$ and
		$\delta^{\left(  2\right)  }$ for different values of $\alpha_{0}$ considering
		$N_{k}=52$ and $\delta_{c}^{2}=\alpha_{0}/3$.}
\begin{tabular}
[c]{|c|c|c|c|}\hline
$\alpha_{0}$ & $\delta_{c}$ & $\delta^{\left(  2\right)  }$ & $\delta^{\left(
	1\right)  }$\\\hline \hline
$10^{-18}$ & $5.8\times10^{-10}$ & $2\times10^{-13}$ & $1.7\times10^{-6}%
$\\\hline
$10^{-5}$ & $1.8\times10^{-3}$ & $1.7992\times10^{-3}$ &
$1.8008\times10^{-3}$\\\hline
$10^{-4}$ & $5.8\times10^{-3}$ & $5.7992\times10^{-3}$ &
$5.8008\times10^{-3}$\\\hline
\end{tabular}
\end{table}

The results in Table \ref{tab:delta-et-inflation} show that for relevant values of $\alpha_{0}$
(typically $\alpha_{0}\gtrsim10^{-5}$), eternal inflation occurs only if
$\chi $, when reaching the attractor line, is very close to $\chi_{c}$.

\bigskip

The discussions carried out previously allow us to conclude that the presence
of an $R^{3}$ term as a correction to the Starobinsky model qualitatively
changes the initial characteristics of inflation. If without the $R^{3}$ term
(almost) any initial condition leads to an inflationary regime, with it, some
kind of fine-tuning is necessary. The accuracy of this fine-tuning is highly
uncertain and depends on the probabilistic structure of the preinflationary
model. Even so, the attractive idea of chaotic inflation \cite{Linde:1983gd,mukhanov} is lost with the inclusion of the $R^{3}$ term regardless of
the value of $\alpha_{0}$.

Another possible scenario that must be taken into account is the one in which
the inflaton field provides different initial conditions for different regions
of space. In this scenario, the existence of an interval in phase space
(however small it may be) that produces an eternal inflation regime becomes
quite relevant. That occurs because in this regime an arbitrarily large volume
is generated. Thus, different regions of space produce different causally
disconnected "universes" achieving the multiverse idea \cite{book:145101}. In this context, it is
reasonable to assume that we live in one of these universes even though the
probability of having initial conditions that lead to eternal inflation is
extremely low. Therefore, in such a scenario, the inclusion of the term
$R^{3}$ in the Starobisnky model does not generate the need for a fine-tuning for physical inflation to take place.

\section{Final comments} \label{sec:final-comments}

In this paper, motivated by the success of the Starobinsky model in providing
a consistent inflationary regime supported by the most current observational
data, we propose an extension to it characterized by adding the higher-order
term $R^{3}$ in the action.

We developed a complete study within the cosmological background. We analyzed
the potential and the phase space of the model so that we observed the
existence of three regions with different dynamics for the scalar field $\chi
$. Such analysis enabled us to qualitatively verify that the introduction of
the $R^{3}$ term in Starobinsky's action restricts the initial conditions that
lead the system to a physical inflation. We described the dynamics of the
scalar field in each of the three regions, namely, the asymptotic region for
$V\left(  \chi\right)  \rightarrow0$ when $\chi\gg1$, the plateau region,
where the slow-roll inflationary regime occurs, and the region where the
reheating phase takes place. While the dynamics along the plateau\emph{
}enabled us to build the quantities with which we can compare the model with
the observations, the reheating phase together with the minimal hypothesis of
the usual couplings between the standard matter fields and gravity allowed us
to obtain a restrictive range for the number of $e$-folds of inflation.
Thus, we duly confronted the model with observational data from Planck,
BICEP3/Keck, and BAO. We saw that as the parameter $\alpha_{0}$ increases, the
region in space $n_{s}\times r$ predicted by the model shifts to the left and
slightly downwards. Furthermore, we can consider that observations from CMB
anisotropies upper limit the value of $\alpha_{0}$ to $10^{-4}$.

It is worth qualitatively commenting on the role that a higher-order $R^{n}$ term with $n>3$
plays in Starobinsky$+R^{n}$ models. Assuming that the higher-order $R^{n}$
term is a small correction to Starobinsky, the author of Ref. \cite{Huang:2013hsb} finds that its associated
potential is similar to the model
discussed in our paper. In this case, we can say that in a classical context the existence of a
maximum critical point with finite $\chi _{c}>0$ in the potential may lead
to a problem of fine-tuning in the initial conditions. On the other hand, the contribution of such higher-order $%
R^{n}$ terms becomes negligible for increasing values of $n$ as shown in Refs. \cite{Huang:2013hsb,Ferrara:2013kca}.

We concluded with the discussion about the limitation that the inclusion of an
$R^{3}$ term imposes on the initial conditions: while in the Starobinsky
model, practically any initial conditions lead the system to an inflationary
regime, the inclusion of such a term causes the need for some kind of fine-tuning. This restricts the initial conditions capable of leading the system to physical inflation, and the concept of chaotic inflation is lost for a
non-negligible value of $\alpha_{0}$. On the other hand, we saw that in a scenario where the initial conditions are such that they produce an eternal inflation regime, the idea of a multiverse is achieved. In this case, the inclusion of the $R^{3}$ term in the Starobinsky model does not require a fine-tuning of the initial conditions for the occurrence of physical inflation.

\section*{Acknowledgments}
G. Rodrigues-da-Silva thanks CAPES/UFRN-RN (Brazil) for financial support, J.
Bezerra-Sobrinho thanks PIBIC CNPq/UFRN-RN (Brazil) for financial support and L. G.
Medeiros acknowledges CNPq-Brazil (Grant No. 308380/2019-3) for partial financial support.

\appendix
\section{Einstein's frame} \label{Ap1}

In this appendix, we rewrite the action%

\begin{equation}
S=\frac{M_{Pl}^{2}}{2}\int d^{4}x\sqrt{-g}\left(  R+\frac{1}{2\kappa_{0}}%
R^{2}+\frac{\alpha_{0}}{3\kappa_{0}^{2}}R^{3}\right)  , \label{S original}%
\end{equation}
in Einstein's frame.

We start by writing a second action in the form%
\begin{equation}
\bar{S}=\frac{M_{Pl}^{2}}{2}\int d^{4}x\sqrt{-g}\left[  \kappa_{0}\left(
\lambda+\frac{1}{2}\lambda^{2}+\frac{\alpha_{0}}{3}\lambda^{3}\right)
+\mu\left(  \frac{R}{\kappa_{0}}-\lambda\right)  \right]  , \label{S barra}%
\end{equation}
where $\mu$ is a Lagrange multiplier. By taking the variation with respect to
$\mu$, we obtain%
\[
\lambda=\frac{R}{\kappa_{0}},
\]
which shows that $S=\bar{S}$. Furthermore, taking the variation with respect
to $\lambda$ we have%
\begin{equation}
\mu=\kappa_{0}\left(  1+\lambda+\alpha_{0}\lambda^{2}\right)  . \label{Aux 1}%
\end{equation}
The next step is inverting Eq. (\ref{Aux 1}) and get $\lambda$ as a function
of $\mu$. By rewriting (\ref{Aux 1}), we get the quadratic equation%
\[
\alpha_{0}\lambda^{2}+\lambda+1-\frac{\mu}{\kappa_{0}}=0,
\]
whose solution is%
\[
\lambda_{\pm}=\frac{-1\pm\sqrt{1-4\alpha_{0}\left(  1-\frac{\mu}{\kappa_{0}%
		}\right)  }}{2\alpha_{0}}.
\]
For the limit $\alpha_{0}\rightarrow0$ to be well defined, we must choose the
positive sign. Thus%
\[
\lambda=\frac{-1+\sqrt{1-4\alpha_{0}\left(  1-\bar{\mu}\right)  }}{2\alpha
	_{0}},
\]
where we define $\bar{\mu}\equiv\mu/\kappa_{0}$. Then we must substitute
$\lambda$ in Eq. (\ref{S barra}). So, using the quadratic equation itself, we
obtain%
\[
\bar{S}=\frac{M_{Pl}^{2}}{2}\int d^{4}x\sqrt{-g}\left\{  \bar{\mu}R-\kappa
_{0}\lambda\left[  \frac{2}{3}\left(  \bar{\mu}-1\right)  -\frac{1}{6}%
\lambda\right]  \right\}  .
\]
By working only with the term that depends on $\lambda$, we can get%
\begin{widetext}
\[
\lambda\left[  \frac{2}{3}\left(  \bar{\mu}-1\right)  -\frac{1}{6}%
\lambda\right]  =-\frac{1}{24\alpha_{0}^{2}}\left(  -1+\sqrt{1-4\alpha
	_{0}\left(  1-\bar{\mu}\right)  }\right)  \left[  -1+8\alpha_{0}\left(
1-\bar{\mu}\right)  +\sqrt{1-4\alpha_{0}\left(  1-\bar{\mu}\right)  }\right]
.
\]
Therefore,%
\[
\bar{S}=\frac{M_{Pl}^{2}}{2}\int d^{4}x\sqrt{-g}\left\{  \bar{\mu}%
R+\frac{\kappa_{0}}{24\alpha_{0}^{2}}\left(  -1+\sqrt{1-4\alpha_{0}\left(
	1-\bar{\mu}\right)  }\right)  \left[  -1+8\alpha_{0}\left(  1-\bar{\mu
}\right)  +\sqrt{1-4\alpha_{0}\left(  1-\bar{\mu}\right)  }\right]  \right\}
.
\]
Then we must carry out the transformations%
\[
\bar{\mu}=e^{\chi}\text{ \ and }\bar{g}_{\mu\nu}=e^{\chi}g_{\mu\nu},
\]
in the action $\bar{S}$. By using the results of Appendix D of Ref.
\cite{book:925650}, concerning the change in Ricci tensor and scalar curvature
due to a conformal transformation, we get%
\begin{align*}
\bar{S} &  =\frac{M_{Pl}^{2}}{2}\int d^{4}x\sqrt{-\bar{g}}\left\{  \bar
{R}-\frac{3}{2}e^{-\chi}g^{\alpha\beta}\partial_{\alpha}\chi\partial_{\beta
}\chi+3e^{-2\chi}\square e^{\chi}\right.  +\\
&  +\left.  \frac{\kappa_{0}e^{-2\chi}}{24\alpha_{0}^{2}}\left(
-1+\sqrt{1-4\alpha_{0}\left(  1-e^{\chi}\right)  }\right)  \left[
-1+8\alpha_{0}\left(  1-e^{\chi}\right)  +\sqrt{1-4\alpha_{0}\left(
	1-e^{\chi}\right)  }\right]  \right\}  .
\end{align*}
Furthermore,%
\[
e^{-2\chi}\square e^{\chi}=e^{-2\chi}\nabla_{\rho}\nabla^{\rho}e^{\chi}%
=\frac{1}{\sqrt{-\bar{g}}}\partial_{\rho
}\left(  g^{\rho\nu}\sqrt{-g}\partial_{\nu}e^{\chi}\right)  ,
\]
is a surface term and it can be neglected. Thus,%
\[
\bar{S}=\frac{M_{Pl}^{2}}{2}\int d^{4}x\sqrt{-\bar{g}}\left\{  \bar{R}%
-\frac{3}{2}\bar{\partial}^{\beta}\chi\bar{\partial}_{\beta}\chi+\frac
{\kappa_{0}e^{-2\chi}}{24\alpha_{0}^{2}}\left(  -1+\sqrt{1-4\alpha_{0}\left(
	1-e^{\chi}\right)  }\right)  \left[  -1+8\alpha_{0}\left(  1-e^{\chi}\right)
+\sqrt{1-4\alpha_{0}\left(  1-e^{\chi}\right)  }\right]  \right\}  .
\]
We can write this expression as%
\[
S\left(  \bar{g}_{\mu\nu},\chi\right)  =\frac{M_{Pl}^{2}}{2}\int d^{4}%
x\sqrt{-\bar{g}}\left[  \bar{R}-3\left(  \frac{1}{2}\bar{\partial}^{\rho}%
\chi\bar{\partial}_{\rho}\chi+V\left(  \chi\right)  \right)  \right]  ,
\]
with
\begin{equation}
V\left(  \chi\right)  =\frac{\kappa_{0}}{72\alpha_{0}^{2}}e^{-2\chi}\left(
1-\sqrt{1-4\alpha_{0}\left(  1-e^{\chi}\right)  }\right)  \left[
-1+8\alpha_{0}\left(  1-e^{\chi}\right)  +\sqrt{1-4\alpha_{0}\left(
	1-e^{\chi}\right)  }\right]  .\nonumber
\end{equation}
It is interesting noting that the Starobinsky limit $\alpha_{0}\rightarrow0$
is well defined. It follows that%
\[
\lim_{\alpha_{0}\rightarrow0}V\left(  \chi\right)  =\frac{\kappa_{0}}%
{6}\left(  1-e^{-\chi}\right)  ^{2},
\]
as expected.
\end{widetext}

\bibliography{references-inflation-R3}
\end{document}